\documentclass[onecolumn,aps,prc,nofootinbib]{revtex4}
\usepackage{amsfonts}
\usepackage{epsfig}
\usepackage{epsfig}
\usepackage{epsfig}
\usepackage{amsmath}
\usepackage{bm}
\usepackage{graphicx}

\newcommand{\beq}{\begin{equation}}
\newcommand{\eeq}{\end{equation}}
\newcommand{\bea}{\vspace{0.25cm}\begin{eqnarray}}
\newcommand{\eea}{\end{eqnarray}}

\newcommand{\nabb}{\mbox{{\boldmath
$\nabla$}}}

\newcommand{\ro}{\mbox{{\boldmath
$\rho$}}}
\newcommand{\ta}{\mbox{{\boldmath
$\tau$}}}
\newcommand{\qb}{\mbox{{\bf
q}}}
\newcommand{\bb}{\mbox{{\bf
b}}}
\newcommand{\pb}{\mbox{{\bf
p}}}

\newcommand{\tb}{\mbox{{\bf
t}}}

\newcommand{\xb}{\mbox{{\bf
x}}}

\newcommand{\Bb}{\mbox{{\bf
B}}}

\newcommand{\Abt}{\mbox{{\bf
A}}_\perp}

\newcommand{\Ab}{\mbox{{\bf
A}}}

\newcommand{\Tb}{\mbox{{\bf
T}}}

\newcommand{\fb}{\mbox{{\bf
f}}}

\newcommand{\vbt}{\mbox{{\bf
v}}_\perp}

\def\lsim{\mathrel{\rlap{\lower4pt\hbox{\hskip1pt$\sim$}}
    \raise1pt\hbox{$<$}}}         %less than or approx. symbol
\def\gsim{\mathrel{\rlap{\lower4pt\hbox{\hskip1pt$\sim$}}
    \raise1pt\hbox{$>$}}}         %greater than or approx. symbol
\long\def\symbolfootnote[#1]#2{\begingroup%
\def\thefootnote{\fnsymbol{footnote}}\footnotemark[#1]\footnotetext[#1]{#2}\endgroup}

\begin{document}
\vspace*{-2cm}
 
\bigskip
%%%%%%%%%%%%%%%%%%%%%%%%%%%%%%%%%%%%%%%%%%%%%%%%%%%%%%%%%%
%%%%%%%%%%%%%%%%%%%%%%%%%%%%%%%%%%%%%%%%%%%%%%%%%%%%%%%%%% 

\begin{center}

\renewcommand{\thefootnote}{\fnsymbol{footnote}}

  {\Large\bf
    Effect of color randomization on $p_T$ broadening of fast
   partons in turbulent quark-gluon plasma
\\
\vspace{.7cm}
  }
\medskip
{\large
  B.G.~Zakharov
  \bigskip
  \\
  }
{\it  
 L.D.~Landau Institute for Theoretical Physics,
        GSP-1, 117940,\\ Kosygina Str. 2, 117334 Moscow, Russia
\vspace{1.7cm}\\}

  {\bf
  Abstract}
\end{center}
{
\baselineskip=9pt
We  analyze  the  effect of the parton color randomization
on $p_T$ broadening
in  the quark-gluon plasma
with turbulent color fields.
We calculate the transport coefficient for a simplified model of
fluctuating color fields in the form of
alternating sequential transverse layers
with homogenous transverse chromomagnetic fields with random orientation in the
SU(3) group and
gaussian distribution in the magnitude.
Our numerical results show that the color randomization
can lead to a sizable reduction of the turbulent contribution
to $\hat{q}$.
The magnitude of the effect grows with increasing
ratio of the electric and magnetic screening masses.

\vspace{.5cm}
}
\section{Introduction}
The experiments on heavy ion collisions at RHIC and the LHC
led to the discovery of quark-gluon plasma (QGP).
The hydrodynamic simulations (for reviews, see, e.g., Refs.
\cite{hydro2,hydro3,Heinz_hydro2}) of $AA$ collisions at the RHIC
and LHC energies show that the QGP produced in $AA$ collisions
flows as almost ideal fluid
 with a very small
shear viscosity to entropy density
ratio $\eta/s$, of the order of the lower bound $\eta/s\sim 1/4\pi$
\cite{KSSeta}.
Hydrodynamical analyses give strong evidence for
the onset of the collective flow/thermalization regime
of the QCD matter at the proper time $\tau_{0}\sim 0.5-1$ fm
\cite{Heinz_hydro2,Heinz_tau}.  
No clear consensus has emerged on whether these facts can be explained within
the paradigm of the weakly coupled QGP.
On the one hand, 
calculations of the shear viscosity in the kinetic theory
with $2\to 2$ parton processes in the leading logarithmic
approximation \cite{AMY_LL} give
$\eta/s\sim 0.6$ (for the QCD coupling constant $g=2$ and $N_F=2.5$),
that is
considerably larger than required for description
of the flow effects observed in $AA$ collisions at RHIC and the LHC.
This contradiction of the kinetic theory with binary collisions
to data on $AA$ collisions was confirmed by direct simulation of
the QGP evolution
in $AA$ collisions at RHIC energies within the Boltzmann equation performed in
\cite{MolnarGyul}.
The inclusion
of the near collinear splitting $1\leftrightarrow2$ processes
\cite{AMY-eta} does not change considerably the ratio $\eta/s$. However, in
\cite{Moore-NLO} it was shown that accounting for the NLO corrections to
the formulas of the effective kinetic approach of \cite{AMY-eta}
can considerably reduce $\eta/s$
\footnote{The large difference between LO and NLO results
  of \cite{Moore-NLO} arises from the large NLO correction
  to $\hat{q}$ obtained in \cite{CH}, which is used in
  the framework of \cite{Moore-NLO} for description of the small angle
parton scattering within the Fokker-Plank approximation.}.
On the other hand, there is indication that 
with the inclusion of the large
angle $2\leftrightarrow 3$ processes
the kinetic approach 
may give $\eta/s\sim 1/4\pi$
\cite{Greiner_eta_3g} without the NLO corrections to the parton
cross sections. Also, the approach of \cite{Greiner_eta_3g}
leads to considerable decrease of the thermalization time 
\cite{Greiner_therm} as compared to
the perturbative ``bottom-up'' scenario \cite{BMSS} based on the
Boltzmann equation with $2\to 2$ and near collinear
$2\leftrightarrow 3$ processes, which does not explain the small
thermalization time. 
However, one should bear in mind that
the analyses
\cite{AMY-eta,MolnarGyul,Moore-NLO,Greiner_eta_3g,Greiner_therm,BMSS}
ignore the possible plasma instabilities that
can appear for anisotropic
initial parton system produced in $AA$ collisions
\cite{RomatStrick1,Arnold_bup1,Mrowc_bup1}.
Possible importance of the QGP instabilities
for $AA$ collisions was first discussed in
\cite{M1}.
One of the important QGP instabilities is
the chromomagnetic
instability  similar to the Weibel instability in
the ordinary plasmas \cite{Weibel}.
The unstable chromomagnetic modes of
the gluon field due to the color Weibel instability
must appear due to strong anisotropy of the initial
parton distribution \cite{BMSS,Arnold_bup1,Mrowc_bup1,Arnold-Moore_anis}
for which the transverse parton momenta are larger than
that along the beam axis.
The generation of the unstable color fields
can accelerate  the thermalization \cite{Arnold-Moore_anis,Arnold2}.
From the point of view of the flow effects in $AA$ collisions, it is
important that interaction of the thermal partons with the random
color fields can reduce the parton mean free path in the QGP, and
consequently can decrease the viscosity of the QGP produced in $AA$
collisions. This mechanism
of generation of a small effective shear viscosity of the QGP in
$AA$ collisions has been addressed in Refs.
\cite{Asakawa1,BMuller1,Muller-Wang-qhat}. Also, in \cite{MY-spin}
it was argued that the turbulent color fields in the QGP can potentially be
important for spin transport in $AA$ collisions.
The effect of the instabilities
on the QGP evolution in $AA$ collisions has attracted much
attention in the literature (for a review, see Refs. \cite{DrLeon,QGPinst}).

The reduction of the QGP shear viscosity in the presence
in the QGP of the turbulent collective color fields
is closely related to the turbulent enhancement of the transport
coefficient $\hat{q}$ \cite{Muller-Wang-qhat}.
The transport coefficient $\hat{q}$
characterizes the mean squared transverse momentum acquired by a fast particle
per unit length in the QGP \cite{BDMPS2}. In terms of $\hat{q}$ one can obtain
the estimate for the typical length, $\lambda$, of degradation
of the momentum for thermal partons in the QGP:
$\lambda=b\langle p\rangle^2/\hat{q}$, where $\langle p\rangle\sim 3T$
is the average parton momentum, and a reasonable choice for
the coefficient $b$ is $b\sim 0.5$.
Then, using the fact that in the kinetic theory
the shear viscosity is approximately \cite{Reif,Gyul-kin}
\beq
\eta\sim \frac{1}{3} n\langle p\rangle
\lambda
\label{eq:10}
\eeq
($n$ is the parton number density) one can obtain
the approximate relation between $\eta$ and $\hat{q}$
\beq
\frac{\eta}{s}\sim \frac{1.5T^3}{\hat{q}}\,,
\label{eq:20}
\eeq
that agrees reasonably  with the estimates given in
\cite{Muller-Wang-qhat,eta-qhat}.

Besides the effect of the turbulent contribution to $\hat{q}$
on the QGP shear viscosity,
it is  also important  for $p_T$ broadening of fast partons
with energy $E\gg T$
produced in hard processes in $AA$ collisions.
Interaction of fast partons with the collective color fields can also lead
to synchrotron-like gluon radiation \cite{Zsync}, which may enhance
jet quenching in $AA$ collisions.
The effect of the collective background color fields on the
$p_T$ broadening of fast partons in an unstable QGP has been
addressed in \cite{Mrow_q1} within the classical approach using
Wong's equations \cite{Wong}.
A quantum calculation of $p_T$ broadening of a fast parton in
random color fields have been performed in \cite{DMuller_q1}
(in the context of the parton $p_T$ broadening in the glasma color fields).
The turbulent contribution to $\hat{q}$ is $\propto g^2\epsilon_f r_c$,
where $\epsilon_f$ is the mean energy density of the turbulent
collective fields,
and $r_c$ is the correlation length of these fields. This result
is natural and quite transparent since $p_T$ broadening arises
due to random transverse momentum kicks with $\Delta p_T^2\sim
g^2 \epsilon_fr_c^2$. 
However, the calculations of \cite{Mrow_q1,DMuller_q1} ignore the
fact that gluon
exchanges between the fast parton and the QGP constituents
can change the color state of the fast parton.
It is clear that such gluon exchanges
will affect the transverse momentum acquired
by the fast parton due to interaction with the collective
color field, since the change of the fast parton color state
changes the Lorentz force. One can expect that this
should reduce the turbulent contribution to $p_T$ broadening,
because the color randomization of the
fast parton due to the gluon exchanges effectively
reduces the correlation length of the Lorentz force in the collective
color fields of the QGP. Study of this effect is of interest
both from the point of view jet quenching and from the point of view
of the turbulent shear viscosity of the QGP
\cite{Asakawa1,BMuller1,Muller-Wang-qhat}.
The purpose of the present paper is to study this effect quantitatively.
We perform calculations for a simplified model of random background
chromomagnetic 
fields with alternate homogeneous field layers.

The paper is organized as follows. In section 2, we give
introductory overview
of calculations of the contributions to $p_T$ broadening from
scattering in turbulent color fields and from scattering on the thermal partons
treating these two mechanisms separately.
We give a new alternative simple derivation
for $p_T$ broadening in the turbulent non-abelian fields.
In  section  3, we develop a formalism for calculation  of
$p_T$ broadening in the QGP with fluctuating background fields
accounting for the effect of the color randomization of fast partons.
In section 4, we present
the results of numerical simulations.
Conclusions  are  contained  in section 5. 

\section{Turbulent and thermal contributions to $p_T$ broadening}
In this section we discuss briefly the quantum formalism for
transverse momentum broadening of fast particles passing through
high-temperature QED and QCD plasmas.
The approach is similar to that used in our previous studies on the propagation
of positronium atoms through a matter \cite{Zpath87} and
on the Landau-Pomeranchuk-Migdal effect in QED and QCD \cite{LCPI1}.

We assume that the plasma is statistically homogeneous and isotropic.
We consider a fast particle with $E\gg m$ with the initial
momentum along the $z$-axis.
At leading order in the
particle energy $E$, the transverse dynamics of fast
particles in an external field can be described in terms
of a 2D Schr\"odinger equation, which governs
the $z$-evolution (on the light-cone $t-z=$const) 
of the transverse wave function \cite{Zpath87,LCPI1}.  
In the transverse Hamiltonian the role of ``mass'' is played by
the particle energy $E$.
The transverse Green function can be written in
the Feynman path integral form \cite{FH}
\beq
K(\ro_2 , z_2 | \ro_1, z_1 ) = 
\int  D \ro \, \exp \left[ i  \int_{z_1}^{z_2} dz
\frac{M(d\ro/dz)^2}{2}
\right]
W(\{\ro\})
\,,
\label{eq:30} 
\eeq
where $M=E$, and
\beq
W(\{\ro\},z_2,z_1)=\exp\left[-ie\int_{z_1}^{z_2}dz\left(A^0-A^3-
  \Abt\cdot\vbt\right)\right]
\label{eq:40}
\eeq
is the Wilson line factor for the trajectory $\ro(z)$
in the external potential $A_\mu(t,\ro(z),z)$ at $t-z=$const.  
In the high energy limit the transverse motion is frozen and
the Green function takes the eikonal form (for an abelian external
field)
\beq
K(\ro_2 , z_2 | \ro_1, z_1 ) = \delta(\ro_2-\ro_1) W(\ro,z_2,z_1)
\,,
\label{eq:50} 
\eeq
where
\beq
W(\ro,z_2,z_1)=\exp\left[-ie\int_{z_1}^{z_2}dz n^\mu A_\mu\right]
=\exp\left[-ie\int_{z_1}^{z_2}dz(A^0-A^3)\right]\,
\label{eq:60}
\eeq
is the lightlike Wilson  line factor with $n^\mu=(1,0,0,1)$
for $\ro=\ro_1=\ro_2$. 

In the present analysis we will discuss $p_T$ broadening in the
eikonal approximation. However, using the path integral representation
(\ref{eq:30}) for the Green functions, one can show that for statistically
transverse uniform medium the calculations of $p_T$ broadening
with accurate Green functions give the same results as the eikonal
approximation \cite{Zpath87,LCPI1}. For this reason, all the
results presented in this paper remain valid beyond the eikonal
approximation.

\subsection{ $p_T$ broadening in turbulent background fields}
\subsubsection{Abelian field}
Let us consider first a charged particle propagating
in a fluctuating electromagnetic field.
For a given external field, the evolution operator for the
transverse density matrix $\rho(\bb,\bb',z)=\phi(\bb,z)\phi^*(\bb',z)$
(hereafter we denote the transverse vectors with bold letters) in the eikonal approximation (corresponding to the diagram of Fig. 1a)
reads
\beq
S(\bb,\bb',z_2|\bb,\bb',z_1)=
W(\bb,z_2,z_1)W^*(\bb',z_2,z_1)=W(\bb,z_2,z_1)W(\bb',z_1,z_2)\,,
\label{eq:70}
\eeq
where $W(\bb,z_2,z_1)$
and $W(\bb',z_1,z_2)$
are the lightlike Wilson line factors
given by (\ref{eq:60}) corresponding to the paths
from $(\bb,z_1)$ to $(\bb,z_2)$ and 
$(\bb',z_2)$ to $(\bb',z_1)$, respectively.
The first equality in (\ref{eq:70}) shows that, formally, the operator $S$ can
also be viewed as the eikonal $S$-matrix for scattering of $e^+e^-$
pair. Below this fact will be used to express the thermal
contribution to $\hat{q}$ in terms of the dipole cross
section $\sigma_{e^+e^-}$ (or $\sigma_{q\bar{q}}$ for the QGP)
as was done in \cite{Zpath87,LCPI1}. But for $p_T$ broadening in
turbulent fields we treat
$S$ as the evolution operator of the density matrix.

For statistically uniform medium, the value of $(\bb+\bb')/2$ is immaterial,
and one can take $\bb=-\ro/2$ and $\bb'=\ro/2$
for the eikonal lightlike lines. We take $z_1=0$ and $z_2=L$.
We will consider the evolution operator as a function of $\ro$ and $L$.
The transverse momentum distribution, $I(\pb)$,
of the final particle (for the initial state with zero transverse momentum)
can be written in terms of the operator $S$ as
\beq
I(\pb)=\Big\langle\Big\langle\langle \rho_{\pb}|S|\rho_0\rangle
\Big\rangle\Big\rangle\,,
\label{eq:80}
\eeq
where $\Big\langle\Big\langle\dots\Big\rangle\Big\rangle$ means
averaging over the ensemble of the background fields,
$\rho_0$ is the density matrix of the initial
state with zero transverse
momentum, and $\rho_{\pb}$ is the density matrix of the final
particle with transverse momentum $\pb$.
We assume that at $z=z_{1,2}$ the magnetic field vanishes, i.e., $\Abt$ is
pure gauge.
The density matrix of a particle
with the physical transverse momentum $\pb$
in the presence of a non-zero pure gauge transverse vector potential
$\Abt$ reads 
\beq
\rho_{\pb}(z_2,\ro)=\frac{1}{2\pi}\exp\left[-i\pb\ro\right]
\cdot W^*(\ro/2,-\ro/2,z_2)\,,
\label{eq:90}
\eeq
where
\beq
W(\bb',\bb,z)=\exp\left[ie\int_{\bb}^{\bb'} d\ta \Abt\right]
\label{eq:100}
\eeq
is the Wilson line factor for the straight transverse path
from $\bb$ to $\bb'$, which
appears due to the Aharonov-Bohm phase shift \cite{AhBohm}.
Note that in the presence of a real magnetic field, when $\Abt$ is not
pure gauge,
the final particle can not be regarded as having a
definite transverse momentum. In this case, $\pb$ can only be determined
with uncertainty of the order of the inverse Larmor radius
(which corresponds to the relative error $\Delta k/k\sim eB/k$).

\begin{figure} %[t]
\begin{center}
\includegraphics[height=2.1cm,angle=0]{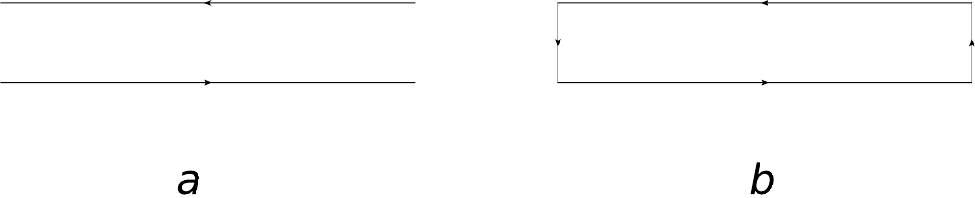}  
\end{center}
\caption[.]{(a) The lightlike Wilson lines
  corresponding to the evolution operator
  for the transverse density matrix of a fast particle
  in the eikonal approximation; (b) the closed Wilson loop
  made of the two lightlike Wilson lines and the transverse
  segments arising from the Aharonov-Bohm phase factors
  in the initial and final density matrices.
}
\end{figure}

Using (\ref{eq:70}), (\ref{eq:80}) and (\ref{eq:90}) one can write the momentum distribution of the
final parton as
\beq
I(\pb)= \frac{1}{(2\pi)^2}\int d\ro
\exp[i\pb\ro]
\Big\langle\Big\langle  W_c(\ro,L)
\Big\rangle\Big\rangle\,,
\label{eq:110}
\eeq
where $W_c$ is the Wilson factor for the closed path as shown in Fig. 1b,
which includes the two straight lightlike lines and the two
transverse segments (corresponding to the Aharonov-Bohm phase factors in the
initial and final density matrices)
\bea
W_c(\ro,L)=
W(-\ro/2,z_1,z_2)W(\ro/2,-\ro/2,z_2)W(\ro/2,z_2,z_1)W(-\ro/2,\ro/2,z_1)
\nonumber\\=
\exp\left[-ie\oint_C dx^\mu \hat{A}_\mu\right]
\label{eq:120}
\eea
with $z_1=0$, $z_2=L$.
The definition of $p_T$ broadening in the QGP in terms of the closed
lightlike Wilson loop was suggested in \cite{UW1,UW2} in
strong-coupling calculations of $\hat{q}$ using AdS/CFT correspondence.
In \cite{UW1,UW2}, it was
assumed that $p_T$ broadening is associated with the long
lightlike Wilson lines, and the effect of two short transverse segments
can be ignored.
However, one should be in mind that in some cases
the whole effect of $p_T$ broadening may come from the transverse Wilson lines.
For instance, this occurs for a fluctuating transverse magnetic
field described by a $z$-dependent transverse vector potential:
$\Bb=\nabla\times \Ab_{\perp}(z)$. In this case
 the lightlike Wilson factors equal unity, and $p_T$ broadening
 arises due to fluctuating Aharonov-Bohm phases in the transverse
 Wilson factors (or in one of them). This occurs even if the transverse Wilson
 lines are located outside the medium.

 We write the averaged $W_c$ in the form
\beq
\Big\langle \Big\langle W_c(\ro,L)\Big\rangle \Big\rangle
=\exp[-P(\ro)]\,.
\label{eq:130}
\eeq
In terms of the function $P(\ro)$, $I(\pb)$ reads
\beq
I(\pb)=\int d\ro \exp\left[i\pb\ro-P(\ro)\right]\,.
\label{eq:140}
\eeq
The value of $\langle \pb^2\rangle$ is sensitive to the behavior
of $P(\ro)$ at small $\rho$.
For particle scattering in the collective turbulent fields we have
$P(\ro)\propto \rho^2$ in the limit $\rho\to 0$.
In this case, from (\ref{eq:140}) one obtains
\beq
\langle \pb^2\rangle=\nabb^2 P(\ro)\Big|_{\rho=0}=
\lim_{\rho\to 0}4P(\rho)/\rho^2\,.
\label{eq:150}
\eeq

The integral over the closed contour $\oint_C dx^\mu A_\mu$
in (\ref{eq:120}) can be transformed in an integral over the surface
spanning it with the help of the Stokes theorem
\beq
\oint_C dx^\mu A_\mu=\frac{1}{2}\int d\sigma^{\mu\nu}F_{\mu\nu}\,.
\label{eq:160}
\eeq
In the limit of small $\ro$,
the surface integral on the right hand side of (\ref{eq:160}) can
be written  as
\beq
\frac{1}{2}\int d\sigma^{\mu\nu}F_{\mu\nu}\approx\int_{0}^{L} dz \ro^j F_{j+}\,,
\label{eq:170}
\eeq
where $j=1,2$ is the transverse index.
Using (\ref{eq:170}) one obtains at small $\ro$
\beq
P(\ro)\approx
\frac{\ro^2e^2}{4}
\int_{0}^{L} dz \int_{0}^{L}dz'
\Big\langle\Big\langle F_{i+}(z)F_{i+}(z')  
\Big\rangle\Big\rangle\,,
\label{eq:180}
\eeq
where, for brevity, for $F_{i+}$ we indicate
only the longitudinal component $z$ of the position four-vector
$x^\mu=(z,0,0,z)$. 
Assuming that
the field correlation radius, $r_c$, is small compared to
the medium thickness $L$,   from (\ref{eq:150}) and (\ref{eq:180}) one obtains
\beq
\langle \pb^2\rangle=2Le^2\int_0^{\infty}dz 
\Big\langle\Big\langle F_{i+}(z)F_{i+}(0)
%f_i(z)f_i(0)  
\Big\rangle\Big\rangle\,.
\label{eq:190}
\eeq
%\beq
This formula can be rewritten as
\beq
\langle \pb^2\rangle=2L\int_0^{\infty}dz 
\Big\langle\Big\langle
%F_{i+}(z)F_{i+}(0)
f_i(z)f_i(0)  
\Big\rangle\Big\rangle\,,
\label{eq:200}
\eeq
where $f_i(z)=-e F_{i+}(z)$
%\label{eq:190}
%\eeq
is the $i$th component
of the Lorentz force
experienced by the charged particle with velocity $v\approx 1$
along $z$ axis. The formulas (\ref{eq:190}) and (\ref{eq:200})
are consistent with the physical picture of $p_T$ broadening as a
random walk in the tranverse
momentum arising from the transverse momentum kicks
with $\langle \Delta\pb^2\rangle\sim e^2\epsilon_f r_c^2$
in scatterings on the domains of the background fields.

\subsubsection{Non-abelian field}
Consider now
$p_T$ broadening in the non-abelian turbulent
collective color fields. It was addressed previously in
\cite{Mrow_q1,DMuller_q1}.
Here we give an alternative simple derivation very similar
to that given above for the abelian case.
In the non-abelian case, the external potential
in the Wilson lines shown in Fig. 1 becomes the color operator
$A_\mu=A_\mu^a\Tb_R^a$, where $\Tb^a_R$ is the color SU(3)
generator for the fast parton. And the abelian formula (\ref{eq:120})
is now replaced by
\beq
\text{Tr} {W}_{c}=
\text{Tr} \left[{\cal{P}}\exp\left(-ig\oint_C dx^\mu
  A_\mu^a\Tb_R^a\right)\right]\,,
\label{eq:210}
\eeq
where ${\cal{P}}$ denotes path ordering of the exponential.
We assume
that the beginning and the end of the path $C$ are at the upper left
corner of the rectangular contour shown in Fig. 1b.
The non-abelian function $P(\ro)$ is defined as
\beq
\frac{1}{d_R}\Big\langle \Big\langle \text{Tr}
{W}_c(\ro,L)\Big\rangle \Big\rangle
=\exp[-P(\ro)]\,,
\label{eq:220}
\eeq
where $d_R$ is the dimension of the color representation for the fast parton
i.e., for SU(3) $d_R=3$ and 8 for quarks and gluons, respectively.

\begin{figure} %[t]
\begin{center}
\includegraphics[height=4.1cm,angle=-90]{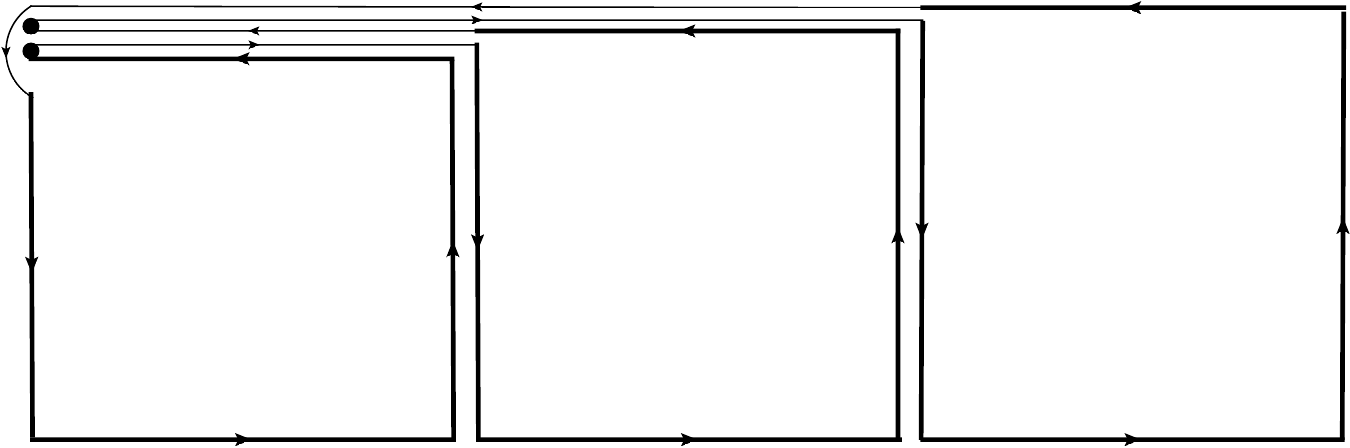}  
\end{center}
\caption[.]
        {A cartoon depicting the deformation of the rectangular
          Wilson loop of Fig. 1b to rewrite 
          the closed non-abelian Wilson line operator in terms of
          a surface integral.
}
\end{figure}

The Wilson line operator $W_c$ can be rewritten
in terms of a surface integral  
with the help of the non-abelian Stokes theorem
\cite{Stokes_Ar,Stokes_Br,Stokes_FGK,Stokes_Si}.
For small $\ro$ this can be done using the deformation of the contour $C$ 
to the one with a sequence of small plaquettes as shown in Fig. 2.
Then, we obtain
\beq
    [{W}_c]_{j}^{i}=
{\cal{P}}_n    \Big[\prod_{n=1}^N {U}(z_n)\Big]_j^{i}\,,
% \Big[\prod_{n=N}^1 {U}_n\Big]_j^{i}\,,    
%\Big[\prod_{n=N^1{U}_n\Big]_j^{i}\,,    
\label{eq:230}    
\eeq
\beq
[U(z_n)]^i_j\approx \delta_j^i-\frac{ig}{2}\Delta \sigma^{\mu\nu}F_{\mu\nu}^a(z_n)
    [\Omega^+_R(z_n)\Tb_R^a\Omega_R(z_n)]^i_j\,,
\label{eq:240}    
\eeq
%where
\beq
\Omega_R(z_n)={\cal{P}}\exp\left[-ig\int_0^{z_n}dz n^\mu A_\mu^a(z)
  \Tb_R^a\right]\,.
\label{eq:250}
\eeq
In (\ref{eq:230})--(\ref{eq:250}) and below we suppress
the time component ($x^0=z$)
and the transverse components ($\xb^{1,2}_\perp=0$) of the position four-vectors
in arguments of $F_{\mu\nu}^a$ and $A_\mu^a$
for notational simplicity.
One can rewrite (\ref{eq:240}) as
\beq
U(z_n)=\hat{1}-\frac{ig}{2}\Delta \sigma^{\mu\nu}
\tilde{F}_{\mu\nu}^a(z_n)\Tb_R^a\,,
\label{eq:260}
\eeq
where  $\tilde{F}_{\mu\nu}=\Omega_A^+F_{\mu\nu}$ is the color rotated
field strength tensor.
Using (\ref{eq:230}), (\ref{eq:260}) and (\ref{eq:210}) one obtains
\beq
\text{Tr} {W}_c\Big|_{\ro\to 0}\approx\text{Tr}
\left[{\cal{P}}\exp\left(-ig\ro^j\int_0^L
  dz \tilde{F}_{j+}^a(z)\Tb_R^a\right)\right]\,.
\label{eq:270}
\eeq
Then,  using the relation
$\text{Tr}(\Tb_R^a\Tb_R^b)=\delta^{ab}d_RC_R/(N_c^2-1)$, from (\ref{eq:270})
and (\ref{eq:150})
we obtain the non-abelian counterparts of (\ref{eq:180}) and (\ref{eq:190})
\beq
P(\ro)\approx \frac{\ro^2g^2C_R}{2(N_c^2-1)}\int_0^L dz_1\int_0^{z_1} dz_2
\Big\langle \Big\langle \tilde{F}_{i+}^a(z_1)\tilde{F}_{i+}^a(z_2) \Big\rangle
\Big\rangle\,,
\label{eq:280}
\eeq
\beq
\langle \pb^2\rangle = \frac{2Lg^2C_R}{N_c^2-1}\int_0^\infty dz
\Big\langle \Big\langle \tilde{F}_{i+}^a(z)\tilde{F}_{i+}^a(0) \Big\rangle\,
\Big\rangle\,.
\label{eq:290}
\eeq

%%%%%%%%%%%%%%%%%%%%%%%%%%%%%%%%%%%%%%%%%%%%%%%%%%%%%%%%%%%%
\subsection{Thermal contribution to $p_T$ broadening}
\subsubsection{Abelian plasma}
Consider now the thermal contribution to $p_T$ broadening for
charged particles in the QED plasma
due to multiple scattering
on the plasma constituents.
The thermal contribution to $p_T$ broadening in the QED plasma
can be treated similarly to the case of ordinary
amorphous materials addressed in \cite{Zpath87}.
The function $P$ corresponding to the product $W W^*$ of the Wilson factors
for the diagram of Fig. 1a, can
be expressed through the Wightman photon field correlator
$\langle\langle A^{\mu}(x)A^{\nu}(y)\rangle\rangle$ as
\beq
P_{th}(\ro)=e^{2}L\int\limits_{-\infty}^{\infty} dz 
[G(z,0_{\perp}z)-G(z,\ro,z)]\,,
\label{eq:300}
\eeq
%where
\beq
G(x-y)= 
n^{\mu}n^{\nu}
{\Large\langle\Large\langle}
A_{\mu}(x)A_{\nu}(y)
{\Large\rangle\Large\rangle}\,.
\label{eq:310}
\eeq
In the approximation of the static Debye screened Coulomb centers
(similar to that of \cite{GW} for the QGP),
neglecting correlations between the plasma constituents, 
the function $P_{th}(\ro)$ given by (\ref{eq:300}) can be written as
\beq
P_{th}^{st}(\ro)= 
\frac{Ln{\sigma}_{e^+e^-}(\rho )}{2}\,,
\label{eq:320}
\eeq
where $n=3\xi(3)T^3/\pi^2$ is the number plasma density (for the
electron-positron plasma with zero chemical potential, and $T\gg m_e$),
and $\sigma_{e^+e^-}(\rho)$ is 
the dipole cross section for scattering of the $e^+e^-$ pair on
the scattering center. In the two-photon exchange approximation
$\sigma_{e^+e^-}(\rho)$ reads
\beq
\sigma_{e^+e^-}(\rho)=\frac{e^{4}}{2\pi^2}\int d\qb
\frac{[1-\exp(i\qb\ro)]}{(\qb^{2}+m_{D}^{2})^{2}}\,\,,
\label{eq:330}
\eeq
where $m_D=eT/\sqrt{3}$ is the Debye mass for the QED plasma \cite{QED_Thoma}.
The formula (\ref{eq:330}) corresponds to the function $P_{th}(\ro)$
\beq
P_{th}^{st}(\ro)=\frac{1}{(2\pi)^2}\int d\qb
[1-\exp(i\qb\ro)]P_{th}^{st}(\qb)
  \label{eq:340}
\eeq
with 
\beq
P_{th}^{st}(\qb)=\frac{Le^43\xi(3)T^3}{\pi^2(q^2+m_D^2)^2}\,.
\label{eq:350}
\eeq

The ratio $C_2(\rho)=\sigma_{e^+e^-}/\rho^2$ (and
$P_{th}^{st}(\rho)/\rho^2$) 
has a logarithmic dependence 
$\propto \ln(\frac{1}{\rho m_D})$
at $\rho\ll 1/m_D$ 
due to the Coulomb tail in 
scattering on the plasma constituents at $q^2\gsim m_D^2$.
Due to the appearance of the Coulomb logarithm,
for the thermal contribution the prescription (\ref{eq:150}) is replaced
by
\beq
\langle \pb^2\rangle\approx \nabb^2 P(\ro)\Big|_{\rho\sim\rho_{min}}\approx
4P(\rho_{min})/\rho^2_{min}\,,
\label{eq:360}
\eeq
where $\rho_{min}\sim 1/q_{max}$, and $q_{max}$ is the maximal
momentum transfer for $2\to 2$ scattering of the fast particle on the plasma
constituents (for a relativistic plasma $q_{max}^2\sim 6ET$).
The transport coefficient $\hat{q}=d\langle \pb^2\rangle/dL$
in terms of the function $C_2$ reads
\beq
\hat{q}_{th}^{st}=2nC_2(\rho\sim \rho_{min})\,.
\label{eq:370}
\eeq
The Coulomb tail leads
to a logarithmic increase of $\hat{q}_{th}$ with the particle energy.
In the static model, from (\ref{eq:340}), (\ref{eq:350})
and (\ref{eq:360}) for $\hat{q}_{th}$  in the
leading-logarithmic
(LL) approximation one obtains 
\beq
\hat{q}_{th}^{st}\approx
\frac{e^43\xi(3)T^3}{4\pi^3}\ln\Big(\frac{6ET}{m_D^2}\Big)\,.
\label{eq:380}
\eeq

In the HTL scheme \cite{HTL}
the $P_{th}(\qb)$ can be written as 
\beq
P_{th}^{HTL}(\qb)=Le^{2}TC(\qb)\,
\label{eq:390}
\eeq
with $C(\qb)=\frac{m_D^2}{q^2(q^2+m_D^2)}$ \cite{AGZ}. This gives
the HTL transport coefficient in the LL approximation 
\beq
\hat{q}_{th}^{HTL}\approx \frac{e^4T^3}{12\pi}\ln\Big(\frac{6ET}{m_D^2}\Big)\,.
\label{eq:400}
\eeq

In the HTL scheme the coefficient of the logarithm in $\hat{q}_{th}$
is smaller by a factor  
of $\pi^2/9\xi(3)\approx 0.912$.
This difference is not surprising since the HTL scheme
is assumed to be valid only at the momenta $\sim eT\ll T$.
In principle, since the Coulomb logarithm comes from the broad region
$6ET\gsim q^2 \gsim m_D^2$, the prediction of the static model
for $\hat{q}_{th}$ should be more accurate
than that of the HTL scheme at $E\gg m_D^2/T$.
Because in this regime the coefficient of the logarithm is controlled
by the number density of the plasma, and this is clearly fulfilled in
the static model.

Note that for the QED plasma the turbulent and thermal
contributions to $\hat{q}$ are additive, since these
two mechanisms do not affect each other.

\subsubsection{Non-abelian plasma}
Calculation of the thermal $p_T$ broadening in the QGP is
very similar that for the QED plasma.
Similarly to QED, we can view the upper line $\leftarrow$ in Fig. 1a
as the Wilson line for antiparticle $\bar{p}$,
since in QCD for a particle $p$ (quark or gluon)
in the color representation $R$,
$-(T^a_R)^*$ equals the color generator $T^a_{\bar{R}}$ for $\bar{p}$. 
Then, similarly to QED, the evolution operator of
the non-abelian density matrix 
can be viewed as $S$-matrix for a fictitious
$p\bar{p}$ state.
This is valid if we restrict ourselves to contributions of one- and
two-gluon exchanges between fast partons and plasma constituents
(as shown in Fig. 3). 
\begin{figure} %[t]
\begin{center}
\includegraphics[height=2.1cm,angle=0]{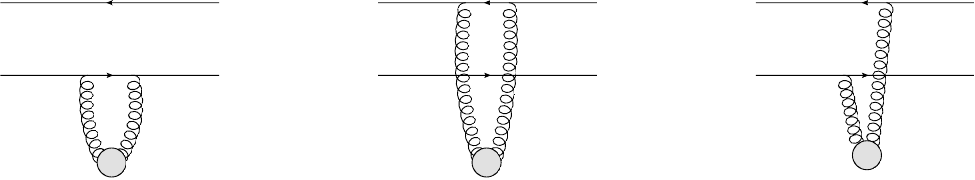}  
\end{center}
\caption[.]
        {Diagrams describing the dipole operator (\ref{eq:410})
          for scattering of $p\bar{p}$ pair on the
          plasma constituent in the double-gluon approximation.
}
\end{figure}

In the approximation of the static Debye screened color centers
\cite{GW}, the two-gluon exchange amplitude of Fig. 3 acts as a
color operator
\beq
\langle p^\beta,\bar{p}^{\bar{\beta}}|\hat{\sigma}(\ro)|
p^\alpha,\bar{p}^{\bar{\alpha}}\rangle=\frac{C_RC_tg^4}{16\pi^2}
\int d\qb \frac{1}{(q^2+m_D^2)^2}\Big[
  \delta_\alpha^\beta\delta^{\bar{\beta}}_{\bar{\alpha}}-(T_R^a)_\alpha^\beta
(T_{\bar{R}}^a)^{\bar{\beta}}_{\bar{\alpha}}\frac{\exp(i\qb\ro)}{C_R}
  \Big]\,,
\label{eq:410}
\eeq
where $m_D=gT\sqrt{1+N_F/6}$ is the Debye mass,
and $C_R$ and $C_t$ are the Casimir 
color operators of the fast parton  and of the QGP constituent. 
This formula accounts for that in Fig. 3,
due to sum over the color indices of the
plasma constituent, the two-gluon $t$-channel states
are color singlets. 
This fact guarantees that, after summing over the color indices for the
initial/final $p\bar{p}$ states, all the intermediate $p\bar{p}$
states between scatterings on different QGP constituents
are color singlets. This allows one in calculating the function $P$
to replace the dipole operator (\ref{eq:410})
by its expectation value between the color singlet $p\bar{p}$ states, i.e.,
by the dipole cross section 
\beq
\sigma_{p\bar{p}}^t(\ro)=\langle \{1\}|\hat{\sigma}(\ro)|
\{1\}\rangle=\frac{C_RC_tg^4}{16\pi^2}
\int d\qb \frac{1}{(q^2+m_D^2)^2}\Big[
1-  \exp(i\qb\ro)
\Big]\,.
\label{eq:420}
    \eeq
Then, one obtains for the QCD counterpart of (\ref{eq:320})
\beq
P_{th}^{st}(\ro)= 
\frac{L\sum_{t=q,g}n_t{\sigma}_{p\bar{p}}^t(\rho )}{2}=
\frac{Ln_q^{eff}{\sigma}_{p\bar{p}}^q(\rho )}{2}
\,,
\label{eq:430}
\eeq
where $n_{q,g}$ are the thermal number densities of quarks and gluons, and
$n_q^{eff}$ is defined as
\beq
n_q^{eff}=n_q+n_gC_A/C_F=\xi(3)[N_F9+C_A16/C_F]T^3/\pi^2\,.
\label{eq:440}
\eeq
Calculating $C_2(\rho_{min})=\sigma_{p\bar{p}}^q(\rho_{min})/\rho_{min}^2$
in the LL approximation one obtains 
the thermal transport coefficient for the static model
\beq
\hat{q}_{th}^{st}\approx
\frac{g^4C_R\xi(3)[C_FN_F9+C_A16]T^3}{32\pi^3}\ln\Big(\frac{6ET}{m_D^2}\Big)\,.
\label{eq:450}
\eeq
In the HTL scheme  we have
\beq
P_{th}^{HTL}(\qb)=Lg^2C_RTC(\qb)\,,
\label{eq:460}
\eeq
and the transport coefficient in the LL approximation is given by
\beq
\hat{q}_{th}^{HTL}\approx
\frac{g^4C_RT^3[1+N_F/6]}{4\pi}\ln\Big(\frac{6ET}{m_D^2}\Big)\,.
\label{eq:470}
\eeq
Comparing (\ref{eq:470}) and (\ref{eq:450}) one sees that in
the LL approximation
the ratio of $\hat{q}_{th}$ for
the HTL scheme to that for the static model is
$\frac{\pi^2(1+N_f/6)}{6\xi(3)(1+N_f/4)}\approx 1.19$ for $N_f=2.5$.

Contrary to the QED plasma for the QGP the contributions
to $p_T$ broadening of the thermal and of the background
field mechanisms are non-additive,
since  $p_T$ broadening due to the background field
is modified because the color exchanges between the fast parton and the medium
constituents lead to change of the Lorentz force experienced by the fast
parton in the background color field.
For this reason, the two mechanisms
of $p_T$ broadening in the QGP
must be treated on an even footing. 

\section{$p_T$ broadening of fast partons in the QGP with simultaneous
  treatment of thermal and background field effects}
In this section we study the non-additivity of $p_T$ broadening in the QGP
corresponding to the background field and to the thermal mechanisms
for a simplified
model of the random background field.
We consider the case of fluctuating layered background color magnetic field
with the transverse layers. We assume that in each layer the magnetic field
is purely transverse and homogeneous.
The fields in different layers are assumed to be uncorrelated.
In this model the layer thickness, $\Delta L$, plays the role of the field
correlation radius in the real turbulent QGP.
The assumption of a purely transverse (and magnetic) field does not seem
to be too restrictive, since
for $p_T$ broadening the crucial quantity is the
transverse Lorentz force acting on the fast parton.
We consider a brick of QGP 
with $N\gg 1$ slabs (i.e., with $L\gg \Delta L$) of fluctuating
homogeneous transverse color magnetic field $\Bb_{a}$. The fields in different
slabs are assumed to be uncorrelated. For the thermal QGP constituents
we use the model of the Debye screened color centers \cite{GW} with
the dipole color operator given by (\ref{eq:410}). Also,
we present the results treating
scattering of the fast parton on the thermal fluctuations
for the HTL scheme with a non-zero magnetic screening mass \cite{AGZ}.
In each slab we take the external vector potential in the form 
\beq
A^{a\mu}=(0,0,0,[\Bb^{a}\times \ro]^{3})=
(0,0,0,\ro\cdot \fb^a)
\,.
\label{eq:480}
\eeq
where $\Bb^{a}$ is the transverse chromomagnetic field, and
$\fb^a$ is the transverse Lorentz force for $g=1$.
We assume that the vector potential is absent at
the initial and final points of the lightlike Wilson lines of Fig. 1a,
and consequently the tranverse Wilson factors can be omitted.
We treat the complex conjugated lightlike Wilson factor of the parton $p$
as that of $\bar{p}$.

Our starting point is the evolution operator  for
the transverse density matrix
for the set $\{f\}=(f_1,...,f_N)$ of the Lorentz forces acting
in the layers $1,\dots,N$, written as
\beq
S(\ro,L,\{f\})=W_p(-\ro/2,L,\{f\})W_{\bar{p}}(\ro/2,L,\{f\})\,.
\label{eq:490}
\eeq
The transverse momentum distribution $I(\pb)$ can be written as
\beq
I(\pb)= \frac{1}{(2\pi)^2}\int d\ro
\exp[i\pb\ro]
\Big\langle\Big\langle  \langle \{1\}|S(\ro,L,\{f\})
|\{1\}\rangle  
\Big\rangle\Big\rangle\,,
\label{eq:500}
\eeq
where $|\{1\}\rangle$ is the color singlet wave function of the
$p\bar{p}$ pair,
and $\Big\langle\Big\langle\dots\Big\rangle\Big\rangle$
means averaging over the ensemble of the fluctuating chromomagnetic
fields and of the positions of the scattering centers. These two averagings
can be performed independently. The averaging over the chromomagnetic
fields is equivalent to averaging over the vectors of
the color Lorentz forces $\fb$ for all the layers.
For a given set $\{f\}$ of the Lorentz forces in the slabs,
the expectation value of the evaluation operator $S$ between the color singlet
initial and final $p\bar{p}$ states can be written as
\bea
\langle \{1\}|{S}(\ro,L,\{f\})|\{1\}\rangle=
\sum_{\Psi_{1},...,\Psi_{N-1}}\langle\{1\}|s(\ro,z_{N+1},z_{N},f_N)|\Psi_{N-1}\rangle
...\nonumber\\
\times
\langle\Psi_2|s(\ro,z_{2},z_{1},f_2)|\Psi_1\rangle
\langle\Psi_1|s(\ro,z_{1},z_{0},f_1)|\{1\}\rangle\,,
\label{eq:510}
\eea
where $z_i=z_1+\Delta L(i-1)$, $\Delta L=L/N$ is the slab thickness,
$f_i$ is the Lorentz force in the slab $i$ (we omit
the transverse space and color indices),
and $s(\ro,z_{i+1},z_i,f_i)$ is the evolution operator for a slab
of the QGP in the interval $(z_i,z_{i+1})$ 
with the Lorentz force $f_i$.
For each of the $i=1,...N$ slabs,
by a proper SU(3) rotation $U_i$,
one can  transform the color Lorentz force vector $f_i^a$ to the one
with components in the Cartan subalgebra
(i.e., with nonzero components only for $a=3$ and $a=8$).
Of course, the necessary matrices $U_i$ may differ for different layers.
Assuming that the background color fields in different slabs are uncorrelated,
the averaging over the background fields can be performed
by independent integrations over the SU(3) rotations
$U_i$ of the Cartan vectors $f_i$ for each of the slabs, and
subsequent averaging over the Cartan vectors $f_i$ in each layer.
Fortunately, one can avoid integrations
over the SU(3) rotations $U_i$. We use the fact that for any color
operator $O$, depending  on the color vector $f$, the color wave function
$|\Psi\rangle$ defined as
\beq
|\Psi\rangle=\int dU O(U f)|\{1\}\rangle
\label{eq:520}
\eeq
can contain only color singlet states.
This means that in calculating the averaged over the chromomagnetic fields
expectation value of the operator $S$, on the right hand side of
(\ref{eq:510})
all the intermediate color states $\Psi_i$
can be replaced by the color singlet state $\{1\}$.
After this replacement, due to the relations
\beq
\langle \{1\}|s(\ro,z_{i+1},z_i,U_if_i)|\{1\}\rangle=
\langle \{1\}|U_is(\ro,z_{i+1},z_i,f_i)U_i^{-1}|\{1\}\rangle=
\langle \{1\}|s(\ro,z_{i+1},z_i,f_i)|\{1\}\rangle\,,
\label{eq:530}
    \eeq
    we can exclude all the SU(3) rotation operators $U_i$.
Thus, we are left only with averaging over the Cartan color vectors
$f_i^a$ with $a=3,8$. Since the integrations over the Cartan vectors
are independent in
different layers, we
    obtain
    \beq
\Big\langle\Big\langle\langle \{1\}|{S}(\ro,L,\{f\})|\{1\}\rangle
\Big\rangle\Big\rangle=
\Big[\Big\langle\Big\langle
\langle \{1\}|s(\ro,\Delta L,0,f)|\{1\}\rangle
\Big\rangle\Big\rangle_f\Big]^N\,,
\label{eq:540}
\eeq
where $\Big\langle\Big\langle\dots
\Big\rangle\Big\rangle_f$ means averaging over the Cartan vectors $f$.
Note that the above formulas remain valid if we include averaging
over the positions of the color centers (corresponding to the thermal
part of $p_T$ broadening) since for the double-gluon exchange
amplitude (shown in Fig. 3) the $t$-channel gluons are
in the color singlet state.
Thus, in our model the problem is reduced to calculation of
the scattering matrix for a single slab of the QGP with a
homogeneous background chromomagnetic
field.
We perform calculations for the Gaussian distributions of the Cartan
components of the Lorentz force color vectors $\fb_a$ with $a=3, 8$ 
\beq
\frac{1}{\pi\sigma^2}\exp\left(-\frac{\fb_a^2}{\sigma^2}\right)\,.
\label{eq:550}
\eeq

%%%%%%%%%%%%%%%%%%%%%%%%%%%%%%%%%%%%%%%%%%%%%%
The interaction of quarks with the Cartan background fields $f_{3,8}$
is diagonal, i.e., it changes only the phase of the $q\bar{q}$ wave function.
But it is not the case for gluons. To diagonalize the interaction of gluons
with the background field, as in \cite{Zsync},  we use the gluon states
having definite
color isospin, $Q_{A}$, and color hypercharge, $Q_{B}$, (we will denote 
the color charge by the two-dimensional vector $Q=(Q_A,Q_B)$).
The diagonal color gluon states,
in terms of the usual gluon vector potential, $G_a$ ($a=1\dots,8$),
read
(the Lorentz indices are omitted)
$X=(G_{1}+iG_{2})/\sqrt{2}$ ($Q=(-1,0)$),
$Y=(G_{4}+iG_{5})/\sqrt{2}$ ($Q=(-1/2,-\sqrt{3}/2)$),
$Z=(G_{6}+iG_{7})/\sqrt{2}$ ($Q=(1/2,-\sqrt{3}/2)$).
The neutral gluons $A=G_{3}$ and $B=G_{3}$ have $Q=(0,0)$,
and do not interact with the background field.
Note that, despite the fact that the initial $p\bar{p}$
state is the color singlet (due to the contraction
of the color indices) after propagation
through the background field the non-singlet components appear,
nevertheless the total color charge $Q_p+Q_{\bar{p}}$ of the $p\bar{p}$ pair 
remains zero.

%%%%%%%%%%%%%%%%%%%%%%%%%%%%%%%%%%%%%%%%%%%%%%%
We assume that the two-gluon
operator (\ref{eq:410}) can be viewed as a local in the
longitudinal coordinate $z$
(this is the standard assumption in models of jet quenching).
In this approximation one can ignore the overlap in the coordinate
$z$ of the background field effects and the two-gluon exchanges
on the $p\bar{p}$ state. We divide the $z$-interval $(0,\Delta L)$
into a linear grid of $M$ cells with width $h=\Delta L/M$.
We assume that the two-gluon
exchanges can only occur in the middle of the cells, i.e., at the transverse
slices $z_i=\Delta L(i-1/2)/M$, $i=1,\dots M$. The background field,
that acts outside of
these slices, changes only the phase of the intermediate $p\bar{p}$ states.
We will denote the color wave functions of the intermediate
$p\bar{p}$ states in the chromomagnetic field as $\psi$. Since
the external field does not change the total color charge
we have only $p\bar{p}$ states with $Q_p+Q_{\bar{p}}=0$.
The phase factor for the intermediate $p\bar{p}$ pair in a color state
$\psi$ 
after propagating
trough the medium a distance $\Delta z$ reads
\beq
\Phi_{\psi}(\Delta z,Q,\ro,f)=\exp
\left[ig \Delta z\ro^i\sum_{a=3,8}Q^af^a_i\right]\,,
\label{eq:560}
\eeq
where $Q^a=(Q_p^a-Q_{\bar{p}}^a)/2=Q_p^a$.

For quarks we have 3 types of the color neutral $q\bar{q}$ intermediate states:
$q^1\bar{q}^1$ ($Q=(1/2,1/2\sqrt{3})$), $q^2\bar{q}^2$ ($Q=(-1/2,1/2\sqrt{3})$),
$q^3\bar{q}^3$ ($Q=(0,-1/\sqrt{3})$). For gluons (in the
basis of charged and neutral gluons) we have 10 types of the color
neutral $g\bar{g}$ states: 6 states made of charged gluons
$X\bar{X}$ ($Q=(-1,0)$), $Y\bar{Y}$ ($Q=(-1/2,-\sqrt{3}/2)$), 
$Z\bar{Z}$ ($Q=(1/2,-\sqrt{3}/2)$), 
$\bar{X}X$ ($Q=(1,0)$), $\bar{Y}Y$ ($Q=(1/2,\sqrt{3}/2)$), 
$\bar{Z}Z$ ($Q=(-1/2,\sqrt{3}/2)$), 
and 4 states with $Q=(0,0)$ made of neutral gluons
$AA$, $BB$, $AB$, $BA$.
So for quarks the operator $s$ is $3\times 3$ matrix, and for gluons
it is $10\times 10$ matrix.

For a given background field (i.e., before averaging over the color fields)
the $s$-matrix element averaged over the thermal states of the QGP
can be written as
\bea
\Big\langle\Big\langle\langle \{1\}|s(\ro,\Delta L,0,\ro,f)|\{1\}\rangle
\Big\rangle\Big\rangle_{th}=
\sum_{\psi_{1},...,\psi_{M+1}}
\langle \{1\}|\psi_{M+1}\rangle
\Phi(\ro,h/2,Q_{M+1},f)\langle \psi_{M+1}|V_{M}(\ro,h)|\psi_{M}\rangle\nonumber\\
\times\Phi(\ro,h,Q_{M},f)\dots
\Phi(\ro,h,Q_{3},f)\langle \psi_3|V_2(\ro,h)|\psi_2\rangle
\Phi(\ro,h,Q_{2},f)\langle \psi_2|V_1(\ro,h)|\psi_1\rangle
\Phi(\ro,h/2,Q_1,f)\langle\psi_1|\{1\}\rangle\,,\,\,\,\,\,\,\,\,\,
\label{eq:570}
\eea
where the operator $V_i$ reads (the index $i$ indicates that the operator
acts at the longitudinal coordinate $z=z_i$)
\beq
V_i(\ro,h)=\hat{1}-\frac{nh}{2}\hat{\sigma}(\ro)\,.
\label{eq:580}
\eeq
The formula (\ref{eq:580}) corresponds to the first order term of the standard
Glauber series, that appears after averaging over positions of the scattering
centers.
One can introduce the total phase factor $\Phi_{tot}(\ro,h,\{Q\},f)$ defined as
the product of the phase factors $\Phi$ for all the intermediate
states $\psi_i$, for the set of the color charges $\{Q\}=(Q_{M+1},\dots,Q_1)$
corresponding to the color charges of the parton $p$ in the
set of the intermediate states 
$\{\psi\}=(\psi_{M+1},\dots,\psi_1)$.
Then, the matrix element of the operator $s$
averaged over the thermal states and over the color Lorentz force $f$ can be
written as
\bea
\Big\langle\Big\langle\langle \{1\}|s(\ro,\Delta L,0,f)|\{1\}\rangle
\Big\rangle\Big\rangle=
\sum_{\psi_{1},...,\psi_{M+1}}
\langle \{1\}|\psi_{M+1}\rangle
\langle \psi_{M+1}|V_M(\ro,h)|\psi_{M}\rangle
\dots\nonumber\\
\times\langle \psi_3|V_2(\ro,h)|\psi_2\rangle
\langle \psi_2|V_1(\ro,h)|\psi_1\rangle\langle\psi_1|\{1\}\rangle
\Big\langle\Big\langle\Phi_{tot}(\ro,h,\{Q\},f)\Big\rangle\Big\rangle_f
\,,\,\,\,\,\,\,\,\,\,
\label{eq:590}
\eea
where the averaged of the color Lorentz force
factor $\Phi_{tot}$ (calculated using the Gaussian distribution (\ref{eq:550}))
reads
\beq
\Big\langle\Big\langle \Phi_{tot}(\ro,h,\{Q\},f)
\Big\rangle\Big\rangle_f=\exp\Big\{-
\frac{\ro^2g^2\sigma^2 h^2}{4}
\sum_{a=3,8}
\Big[\sum_{k=1}^{M+1}\eta_kQ_k^a\Big]^2
\Big\}\,
\label{eq:600}
\eeq
with $\eta_k=1/2$ for $k=1$ and $M+1$, and $\eta_k=1$ for $1<k<M+1$.

The $\langle \pb^2\rangle $ for one slab of the QGP with the background field
in terms of $s$ reads
\beq
\langle \pb^2\rangle=-4\frac{d}{d\rho^2}
\ln\Big[\Big\langle\Big\langle\langle \{1\}|s(\ro,\Delta L,0,f)|\{1\}\rangle
\Big\rangle\Big\rangle\Big]
\Big|_{\rho\sim 1/q_{max}}\,.
\label{eq:610}
  \eeq

  To calculate $\langle \pb^2\rangle$ uing (\ref{eq:610}) we write
  the operator $V$ as
  \beq
  V(\ro,h)= V(0,h)+v(\ro,h)
\label{eq:620}
  \eeq
  with $v(\ro,h)=V(\ro,h)-V(0,h)$.  
Using (\ref{eq:410}) and (\ref{eq:580}) we obtain at small $\ro$
  \beq
  V^{\beta\bar{\beta}}_{\alpha\bar{\alpha}}(0,h)=
  \delta^{\beta}_{\alpha}\delta^{\bar{\beta}}_{\bar{\alpha}}
  -\frac{nhC_RC_tg^4}{32\pi m_D^2}
  \Big[\delta^{\beta}_{\alpha}\delta^{\bar{\beta}}_{\bar{\alpha}}-
\frac{(T^a_R)^{\beta}_{\alpha}
  (T^a_{\bar{R}})^{\bar{\beta}}_{\bar{\alpha}}}{C_R}\Big]\,,
\label{eq:630}  
  \eeq
  \beq
  v^{\beta\bar{\beta}}_{\alpha\bar{\alpha}}(\ro,h)\approx   
-\frac{\rho^2h n_q^{eff}C_2(\rho_{min})}{2C_R}(T^a_R)^{\beta}_{\alpha}
(T^a_{\bar{R}})^{\bar{\beta}}_{\bar{\alpha}}
=-\frac{\rho^2h\hat{q}_R}{4C_R}(T^a_R)^{\beta}_{\alpha}
(T^a_{\bar{R}})^{\bar{\beta}}_{\bar{\alpha}}\,.
\label{eq:640}
\eeq
In (\ref{eq:640}) we used that $\hat{q}=2nC_2$. Note that $V(\ro=0,h)\ne \hat{1}$ because the operator $\hat{\sigma}(\ro)$
given by (\ref{eq:410}) does not vanish at $\ro=0$ (only its expectation value
between color singlet states vanishes at $\ro=0$).
Note that the $V(0,h)$ 
can be rewritten in terms
of the parton
mean free path in the QGP $\lambda_{mfp}=1/n\sigma_p$
as
  \beq
  V^{\beta\bar{\beta}}_{\alpha\bar{\alpha}}(0,h)=
  \delta^{\beta}_{\alpha}\delta^{\bar{\beta}}_{\bar{\alpha}}
  -\frac{h}{\lambda_{mfp}}
  \Big[\delta^{\beta}_{\alpha}\delta^{\bar{\beta}}_{\bar{\alpha}}-
\frac{(T^a_R)^{\beta}_{\alpha}
  (T^a_{\bar{R}})^{\bar{\beta}}_{\bar{\alpha}}}{C_R}\Big]\,,
\label{eq:650}
  \eeq
with
  \beq
\lambda_{mfp}^{-1}=n\int_0^{q^2_{max}} dq^2 \frac{d\sigma_p}{dq^2}\approx
\frac{nC_RC_tg^4}{32\pi m_D^2}\,.
\label{eq:660}
    \eeq
    The  $\lambda_{mfp}$ characterizes the length of the color randomization
    of the parton $p$ in the QGP \cite{GS}. Note that $\lambda_{mfp}$
    defined by (\ref{eq:660}) is smaller that the mean free path $\lambda$ in the formula
of the kinetic theory \cite{Reif,Gyul-kin} for $\eta$ (\ref{eq:10}).
The latter
    is related to 
  the transport cross section $\lambda^{-1}=n\sigma_p^{tr}$ with
    $\sigma_p^{tr}=\int d\sigma_psin^2\theta
    \approx (4/s)\int d\sigma_p q^2$,
    which is smaller than the total cross section entering (\ref{eq:660}).
    Note that the HTL
 counterpart of (\ref{eq:660}), defined by
  \beq
  \lambda_{mfp}^{-1}\Big|_{HTL}\approx
\frac{g^2C_RT}{4\pi}
\int_0^{q^2_{max}} dq^2 C(q^2)
\label{eq:670}
    \eeq
    gives logarithmically divergent $\lambda_{mfp}^{-1}$
    for the standard HTL formula for $C(q^2)$ derived in \cite{AGZ}
    \beq
    C(q^2)=\frac{1}{q^2}-\frac{1}{q^2+m_D^2}\,.
    \label{eq:680}
    \eeq
    This occurs due to the $1/q^2$ term in (\ref{eq:680}), which comes from
    the transverse part of the gluon polarization tensor $\Pi_T$,
    and is related to the absence of screening for the transverse
    static gluons, i.e., due to zero magnetic screening mass,
    defined in the HTL scheme as $m^2_{M}=\text{Re}\Pi_T(0)=0$ \cite{AGZ}.
    In general, it is invalid \cite{Gross_QCD_RMP}.
    In \cite{AGZ} it was suggested
    to account for the nonzero magnetic mass in the HTL calculation
    of the factor $C$
    by using $\text{Re}\Pi_T(0)=m^2_M$ with nonzero $m_M$. This
    leads to
    \beq
    C(q^2)=\frac{1}{q^2+m_M^2}-\frac{1}{q^2+m_D^2}\,.
    \label{eq:690}
    \eeq
    Substituting this into (\ref{eq:670}), we obtain
  \beq
  \lambda_{mfp}^{-1}\Big|_{HTL,m_M\ne 0}\approx
  \frac{g^2C_RT}{4\pi}\ln\Big(\frac{m_D^2}{m_M^2}\Big)\,.
  \label{eq:700}
    \eeq
    Lattice calculations of \cite{m_mag1,m_mag2} give $m_D/m_M\sim 2-3$.
    This can lead to a sizeable logarithmic increase of $\lambda_{mfp}$
    for the HTL scheme with magnetic mass as compared to the model
    with  the static color centers \cite{GW}. Note that the sensitivity of
    the color
    relaxation time (and conductivity) to the ratio $m_D/m_M$
    was discussed long ago in \cite{GS}.
     
    From (\ref{eq:610}) using (\ref{eq:590}), (\ref{eq:600}),
    (\ref{eq:630}) and (\ref{eq:640}) we obtain for one slab
\beq
\langle \pb^2\rangle
=\langle \pb^2\rangle_{th}+\langle \pb^2\rangle_{f}\,,
\label{eq:710}
  \eeq
  where $\langle \pb^2\rangle_{th,f}$ correspond to $p_T$ broadening
due to rescatterings on the 
thermal QGP constituents and scattering in the background turbulent field
affected by the color randomization due to particle rescatterings on
  the thermal partons.
The thermal contribution, which comes from
$v$, reads
\beq
\langle \pb^2\rangle_{th}=
-\frac{Mh\hat{q}_R}{C_R}  \{1\}_{\beta}^{\bar{\beta}}(T^a_R)^{\beta}_{\alpha}
(T^a_{\bar{R}})^{\bar{\beta}}_{\bar{\alpha}}\{1\}_{\alpha}^{\bar{\alpha}}
=\Delta L\hat{q_R}\,.
\label{eq:720}
\eeq
To obtain (\ref{eq:720}), we used that in (\ref{eq:590}),
without the phase factor
$\Phi_{tot}$, all the intermediate projectors
$|\psi_k\rangle\langle \psi_k|$ can be replaced by
$|\{1\}\rangle\langle \{1\}|$. In this case, the right hand side of
(\ref{eq:590})
in the limit of small $\rho$ becomes
\beq
\langle \{1\}|V(0,h)+v(\ro,h)|\{1\}\rangle^M\approx 1+M
\langle \{1\}|v(\ro,h)|\{1\}\rangle\,,
\label{eq:730}
\eeq
that, after substitution into (\ref{eq:610}), leads to (\ref{eq:720})
(which agrees with calculations of the thermal $p_T$
given in section 2, as it must be).
The $\langle \pb^2\rangle_{f}$ term in (\ref{eq:710})
comes from the Taylor expansion of the phase factor $\Phi_{tot}$ and the matrix
elements of $V(0,h)$ in (\ref{eq:590}). Using (\ref{eq:590}), (\ref{eq:600})
and (\ref{eq:610}) one can obtain
\bea
\langle \pb^2\rangle_{f}=
h^2g^2\sigma^2
\sum_{\psi_{1},...,\psi_{M+1}}
\langle \{1\}|\psi_{M+1}\rangle
\langle \psi_{M+1}|V(0,h)|\psi_{M}\rangle
\dots\nonumber\\
\times\langle \psi_3|V(0,h)|\psi_2\rangle
\langle \psi_2|V(0,h)|\psi_1\rangle\langle\psi_1|\{1\}\rangle
\times\sum_{a=3,8}\Big[\sum_{k=1}^{M+1}\eta_kQ_k^a\Big]^2\,.
\label{eq:740}
\eea
The background field contribution to the
transport coefficient is $\hat{q}_{R}^f=\langle p_T^2\rangle_{f}/\Delta L$.
The formula (\ref{eq:740}), together with (\ref{eq:630}),
can be used for numerical calculation of
the turbulent contribution to the transport coefficient.
Note that for quarks summing over the color index $a$ in the second
term inside the square brackets in (\ref{eq:630}) can be performed analytically
with the help of the Fierz identity for triplet color
color generators:
$(\tb^a)^l_m(\tb^a)^i_k=(1/2)\delta^i_m\delta^l_k-(1/2N_c)\delta^i_k\delta^l_m$.
For gluon color generators we performed summing over $a$ in
(\ref{eq:630}) numerically.

In the absence of rescatterings we have $V(0,h)=1$
(see Eq. (\ref{eq:630}), and  (\ref{eq:740})
is reduced to
\beq
\langle \pb^2\rangle_{f}=h^2M^2g^2\sigma^2
\sum_{a=3,8}
\langle Q^{a2}\rangle_{\{1\}}\,.
\label{eq:750}
\eeq
From this relation, using
$\langle \{1\}|Q_3^2+Q_8^2|\{1\}\rangle=2C_R/(N_c^2-1)$,
we obtain in our model without the thermal contribution
\beq
\hat{q}_R=\frac{2C_R\Delta L g^2\sigma^2}{N_c^2-1}\,.
\label{eq:760}
\eeq
This prediction agrees with formula (\ref{eq:290})
because in our model
\beq
\Big\langle \Big\langle \tilde{F}_{i+}^a(z)\tilde{F}_{i+}^a(0) \Big\rangle
\Big\rangle=\Big\langle \Big\langle \tilde{F}_{i+}^a(0)\tilde{F}_{i+}^a(0)
\Big\rangle
\Big\rangle \cdot F(z)
\label{eq:770}
\eeq
with
$
\Big\langle \Big\langle \tilde{F}_{i+}^a(0)\tilde{F}_{i+}^a(0)
\Big\rangle
\Big\rangle=2\sigma^2\,, 
$
and 
\beq
F(z) = \begin{cases}
1-|z|/\Delta L  & \mbox{if } |z| < \Delta L\;,\\
0 &  \mbox{if } |z| > \Delta L\;. \\
\end{cases}
\label{eq:780}
\eeq

It is worth noting that the formulas (\ref{eq:770}) and (\ref{eq:780})
show that our simplified model of the random background
fields nevertheless leads to a quite reasonable
form of the field strength correlation function.
For this reason, one can expect that the use of the multilayer homogeneous
fields cannot greatly distort the results.

Note, finally, that although in our model we use a specific geometry
for the color
exchanges,
physically, it is evident that it cannot be crucial
for our predictions. Because the only important quantity for
the magnitude of the effect of the color randomization on $p_T$
broadening is the ratio $\Delta L/\lambda_{mfp}$, which
is clearly insensitive to the specific choice of the $z$-distribution of
the color exchanges of the fast parton in a slab.

\section{Numerical results}
\begin{figure} %[t]
\begin{center}
\includegraphics[height=3.7cm,angle=0]{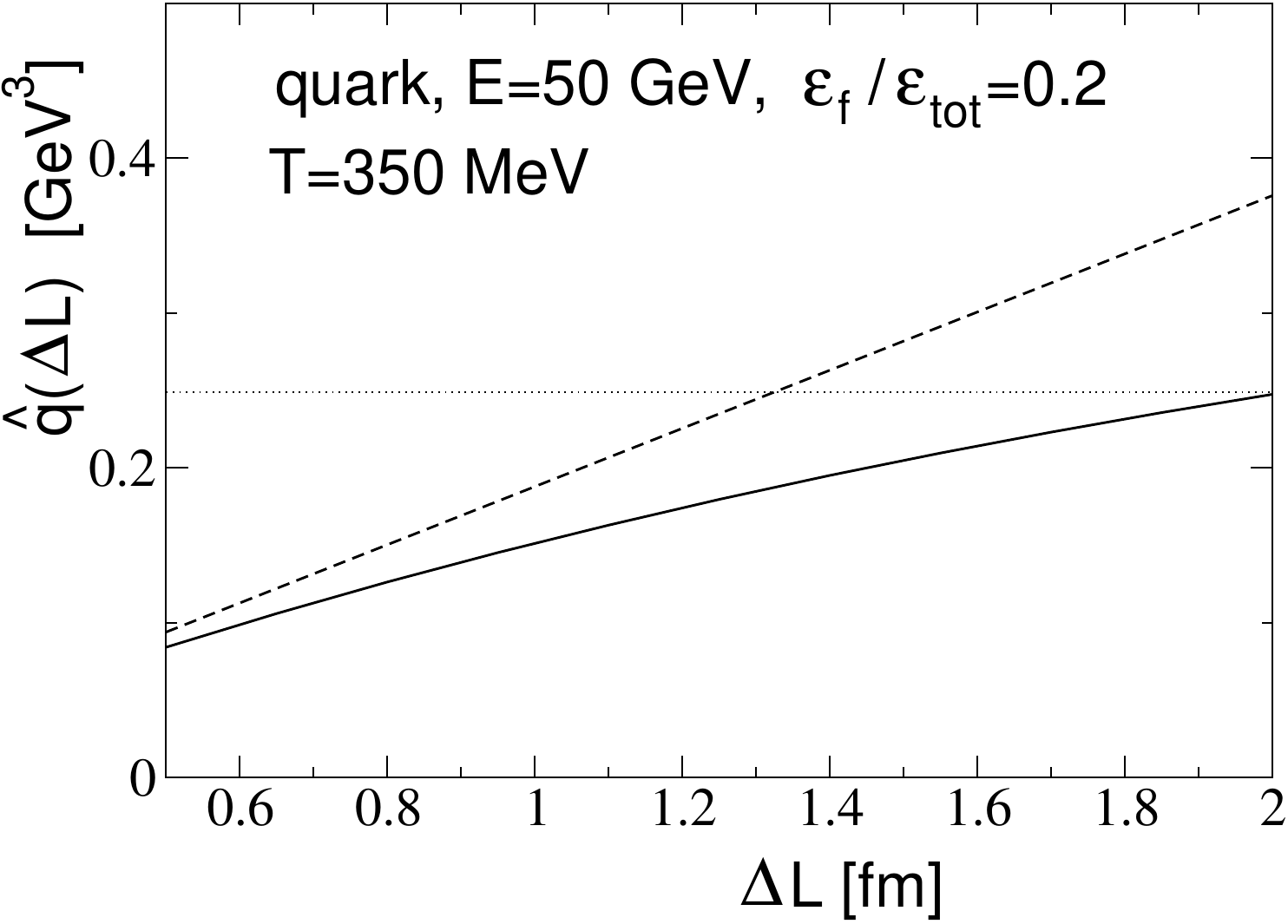}  
\includegraphics[height=3.7cm,angle=0]{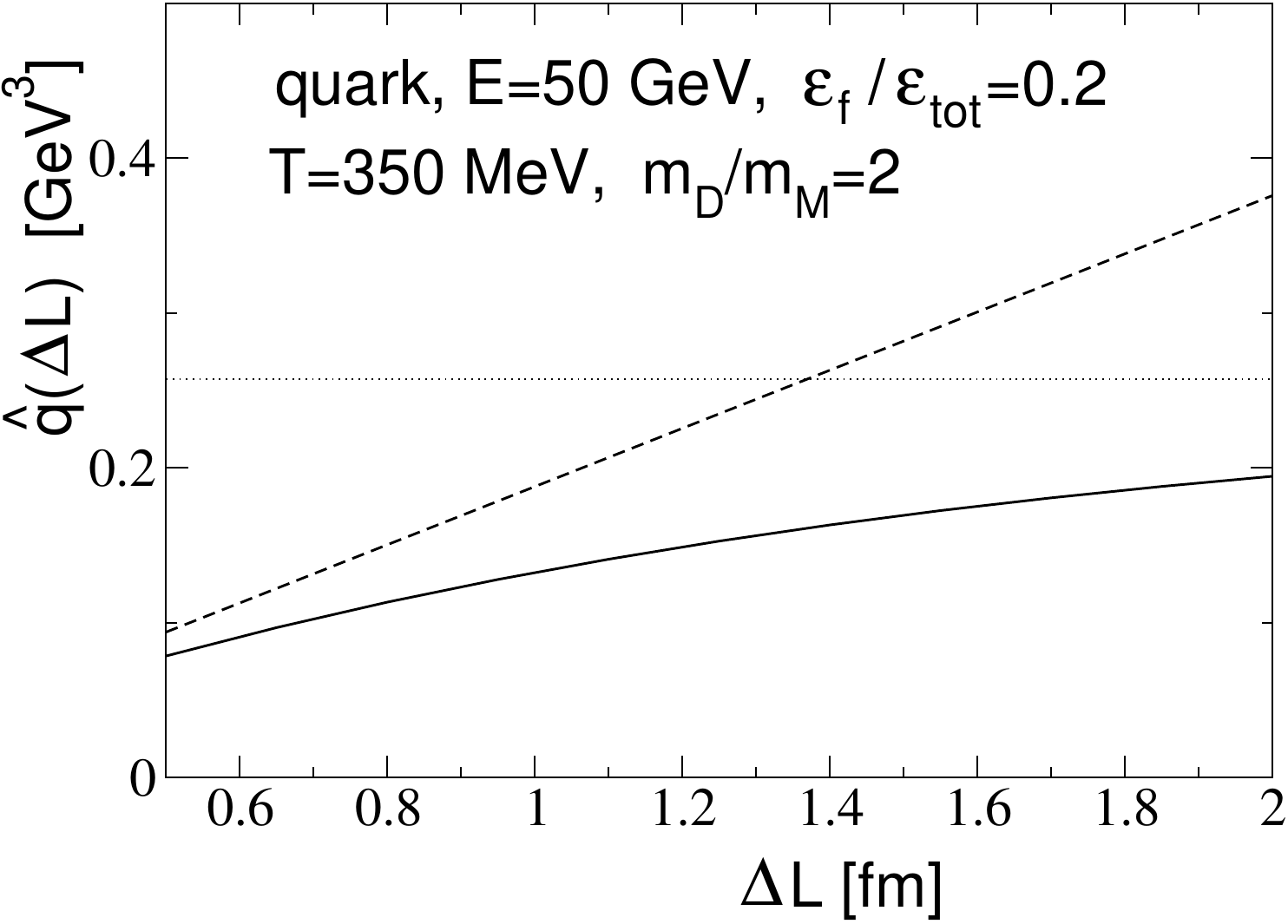}  
\includegraphics[height=3.7cm,angle=0]{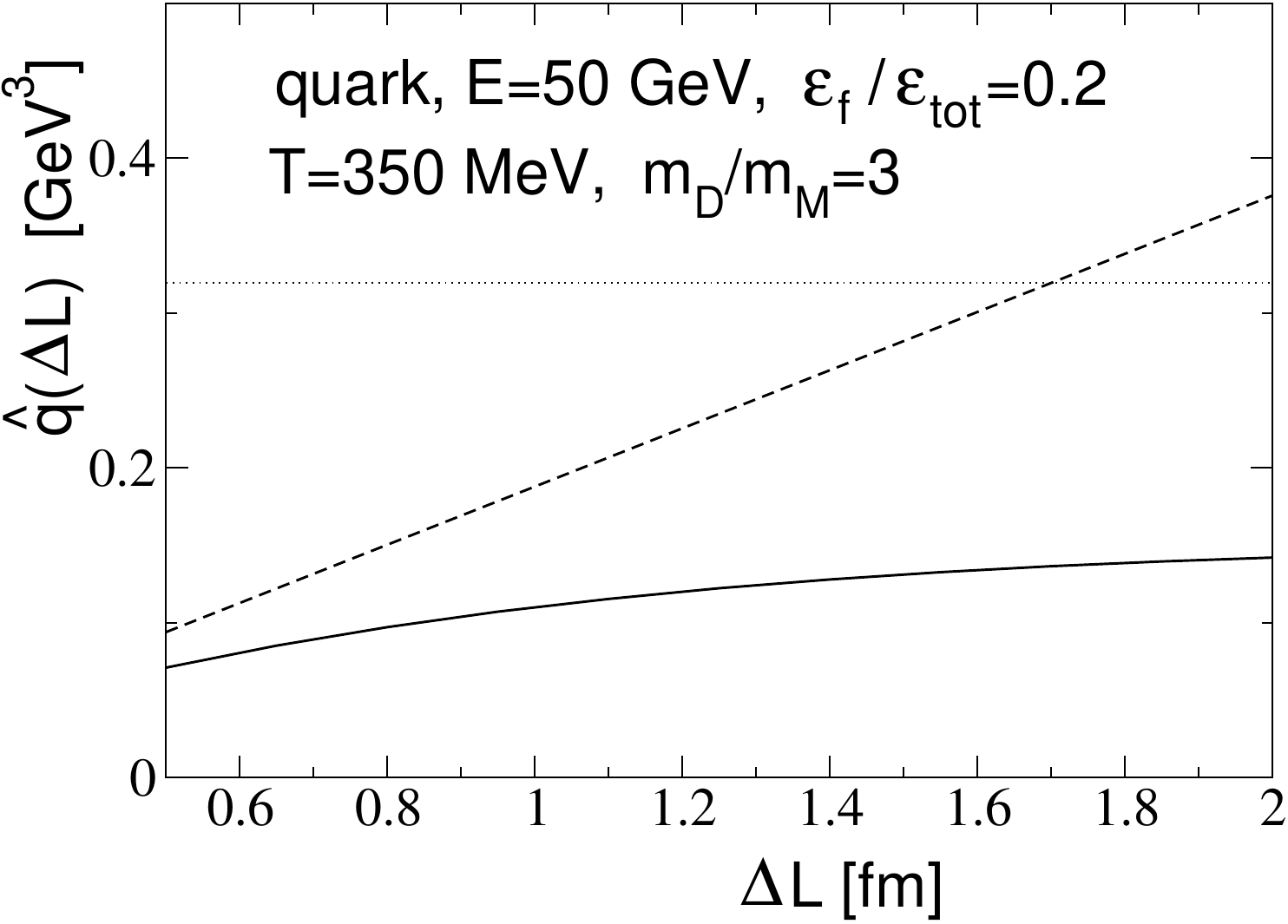}  
\end{center}
\caption[.]{The transport coefficient $\hat{q}$ for quark with
  energy $E=50$ GeV
  in the QGP with $T=350$ MeV, $\epsilon_f/\epsilon_{tot}=0.2$ as a function
  of $\Delta L$
  for the static model (left) and for the HTL scheme with $m_D/m_M=2$ (middle)
  and $m_D/m_M=3$ (right). Solid line: the turbulent contribution to $\hat{q}$
  obtained accounting for the color randomization of the fast parton;
  dashed line: the turbulent contribution without the color randomization
  of the fast parton; dashed line: the thermal contribution.
}
\end{figure}
\begin{figure} %[t]
\begin{center}
\includegraphics[height=3.7cm,angle=0]{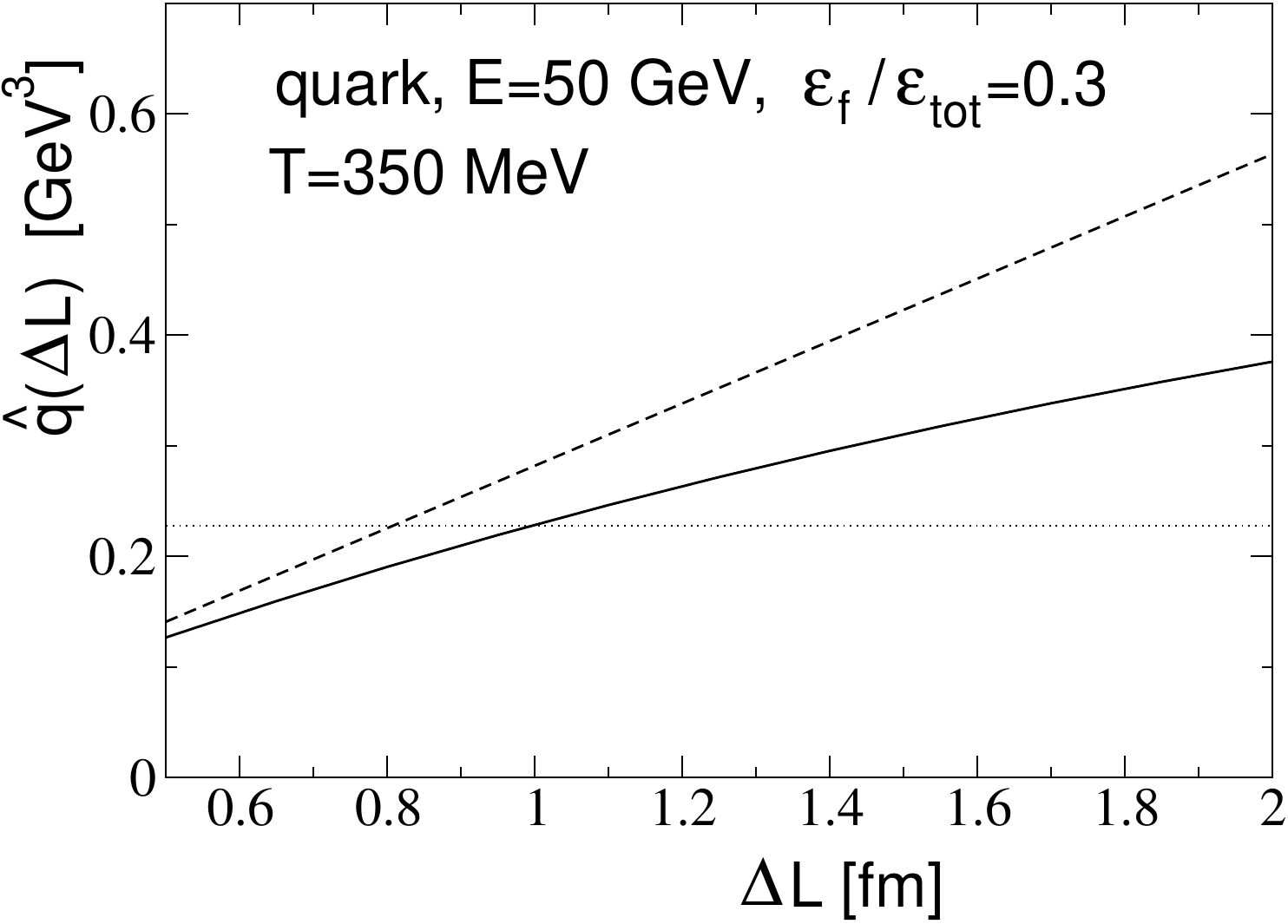}  
\includegraphics[height=3.7cm,angle=0]{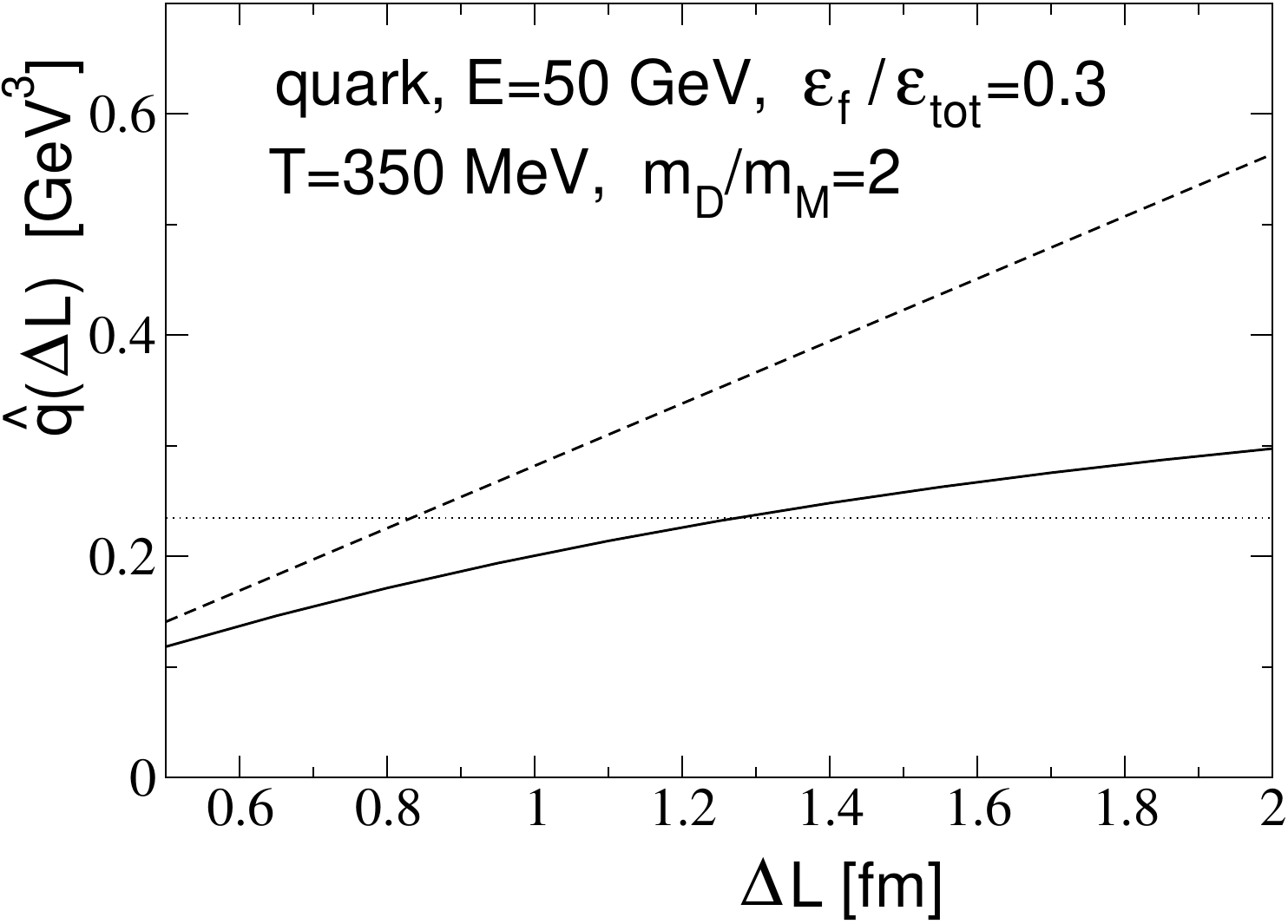}  
\includegraphics[height=3.7cm,angle=0]{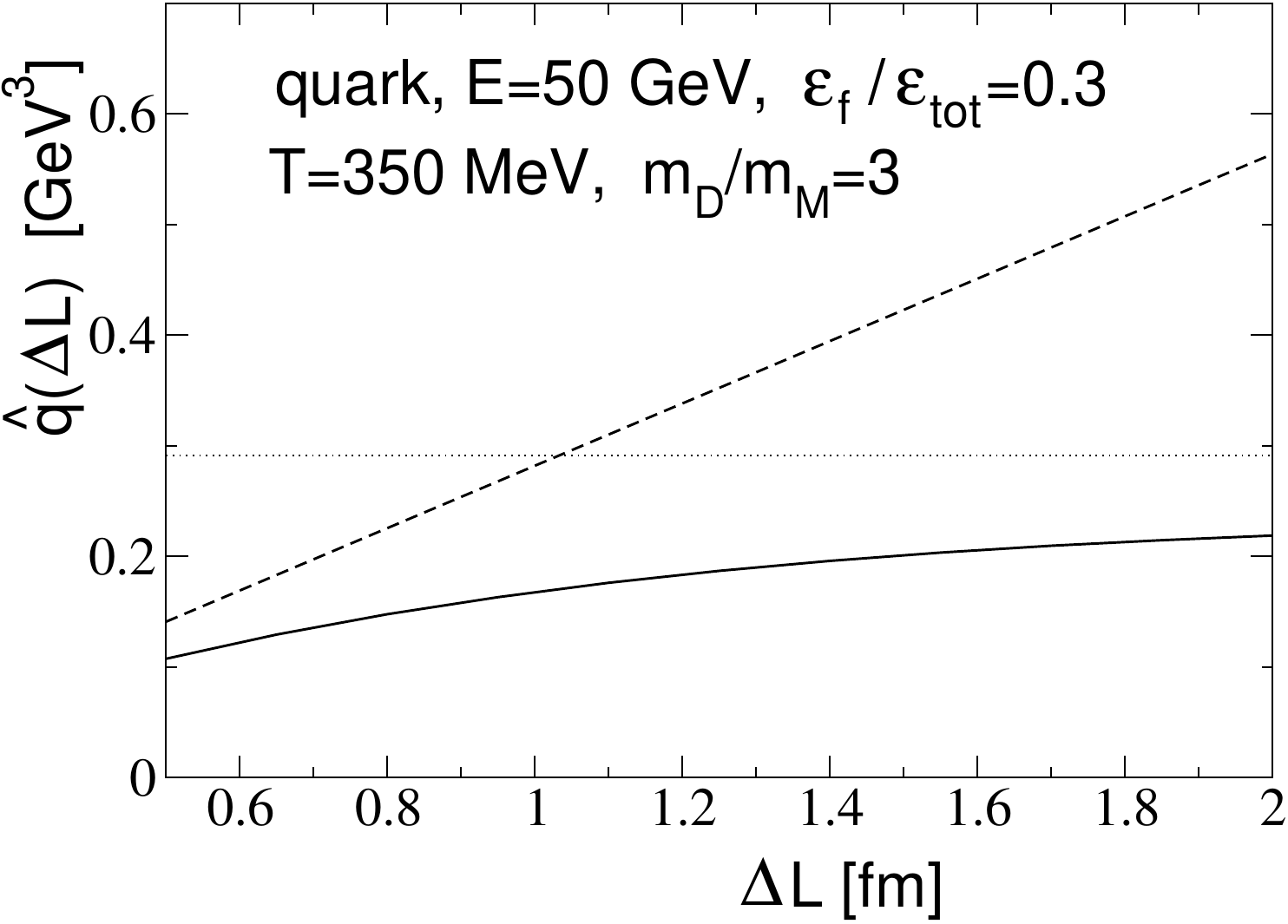}  
\end{center}
\caption[.]{
  The same as in Fig. 4 for $\epsilon_f/\epsilon_{tot}=0.3$.
}
\end{figure}
\begin{figure} %[t]
\begin{center}
\includegraphics[height=3.7cm,angle=0]{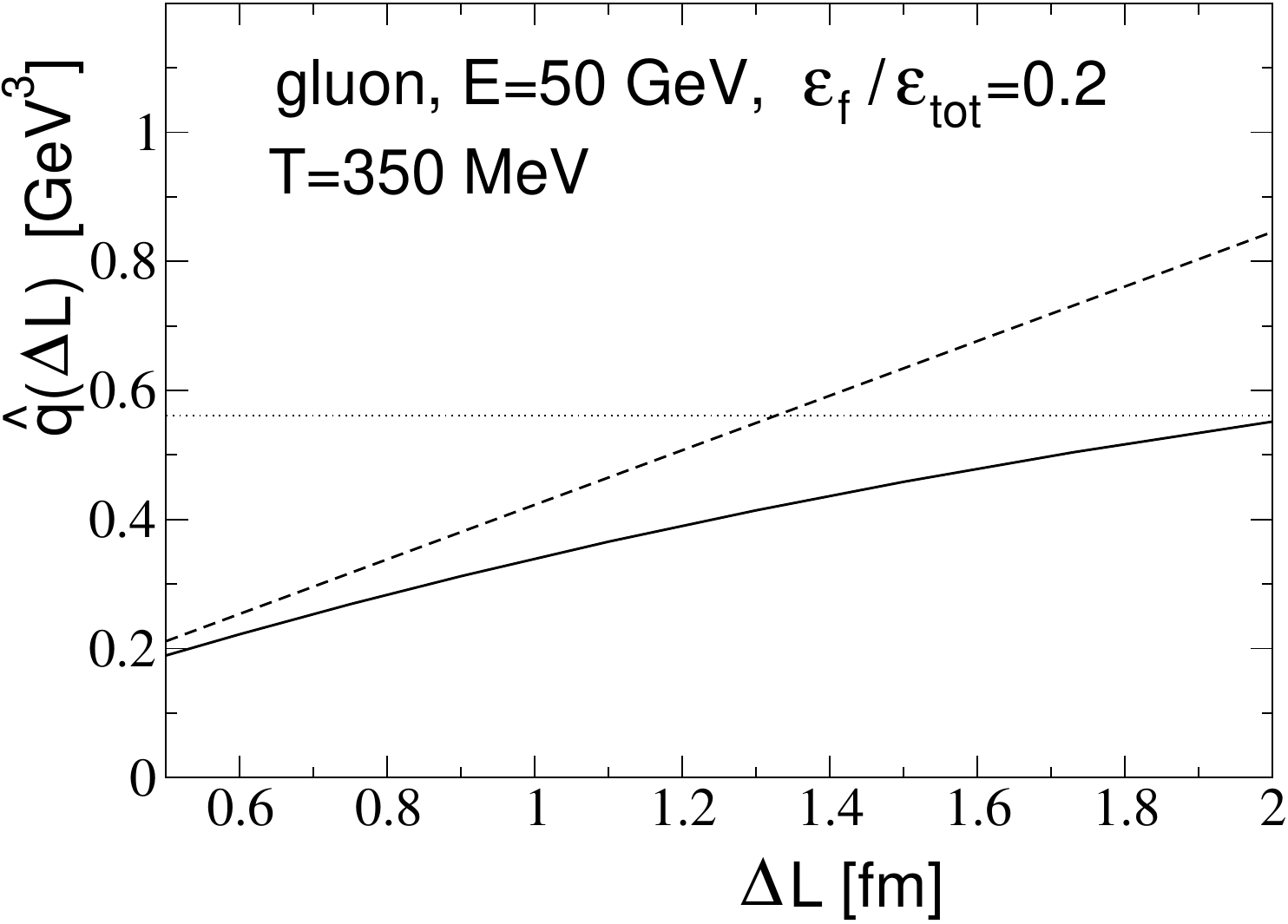}  
\includegraphics[height=3.7cm,angle=0]{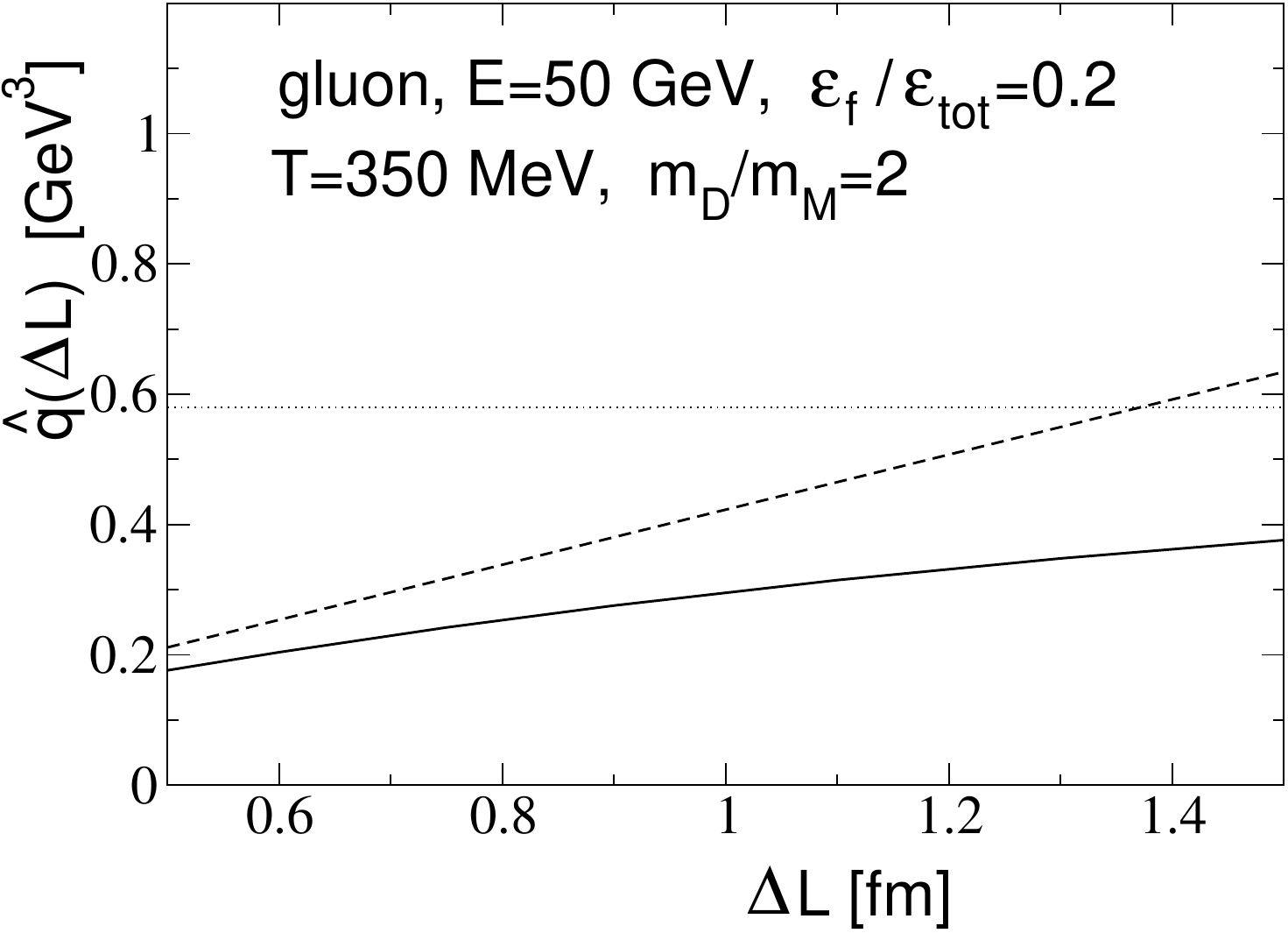}  
\includegraphics[height=3.7cm,angle=0]{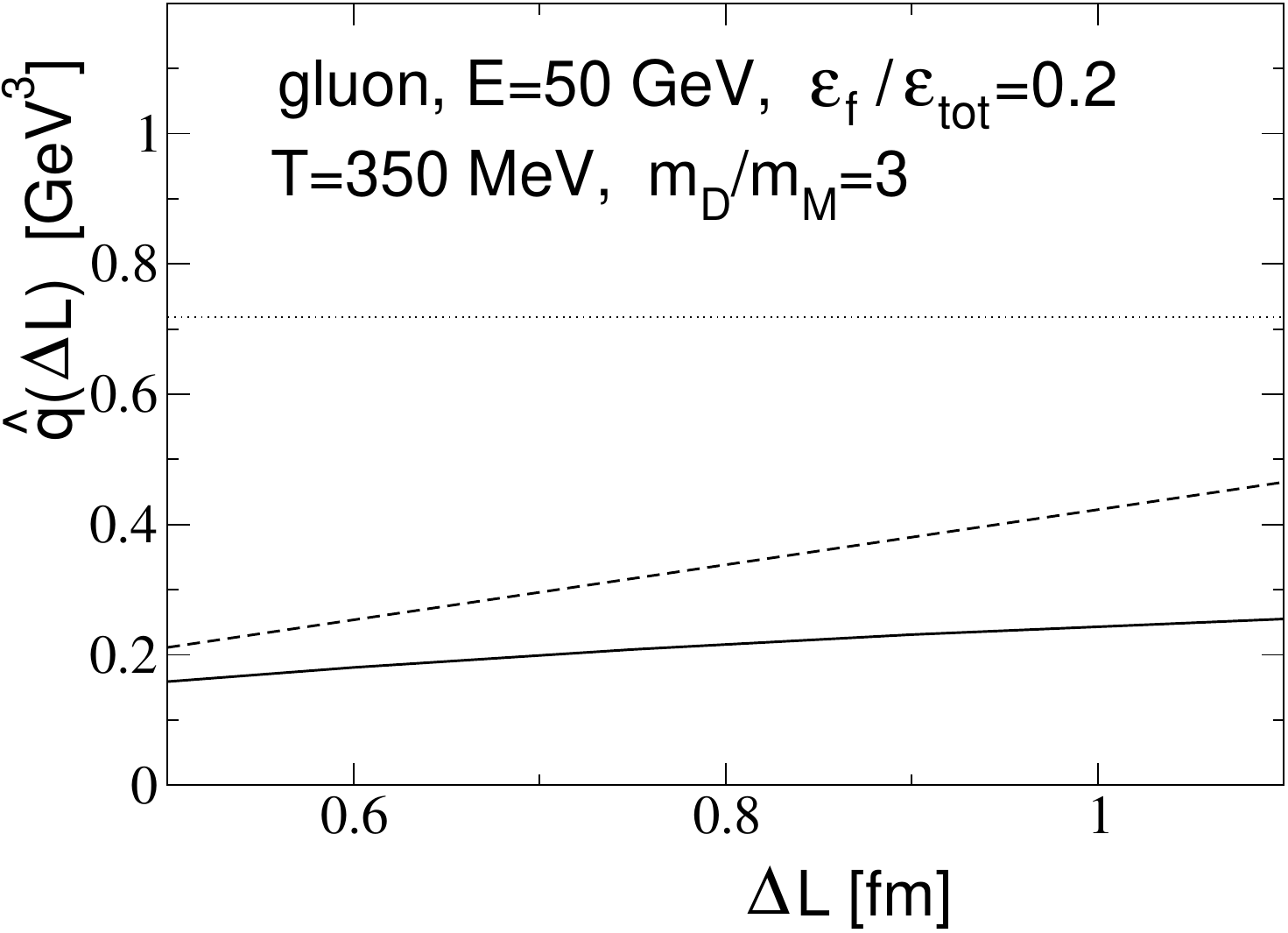}  
\end{center}
\caption[.]{
The same as in Fig. 4 for gluon.
}
\end{figure}
\begin{figure} %[t]
\begin{center}
\includegraphics[height=3.7cm,angle=0]{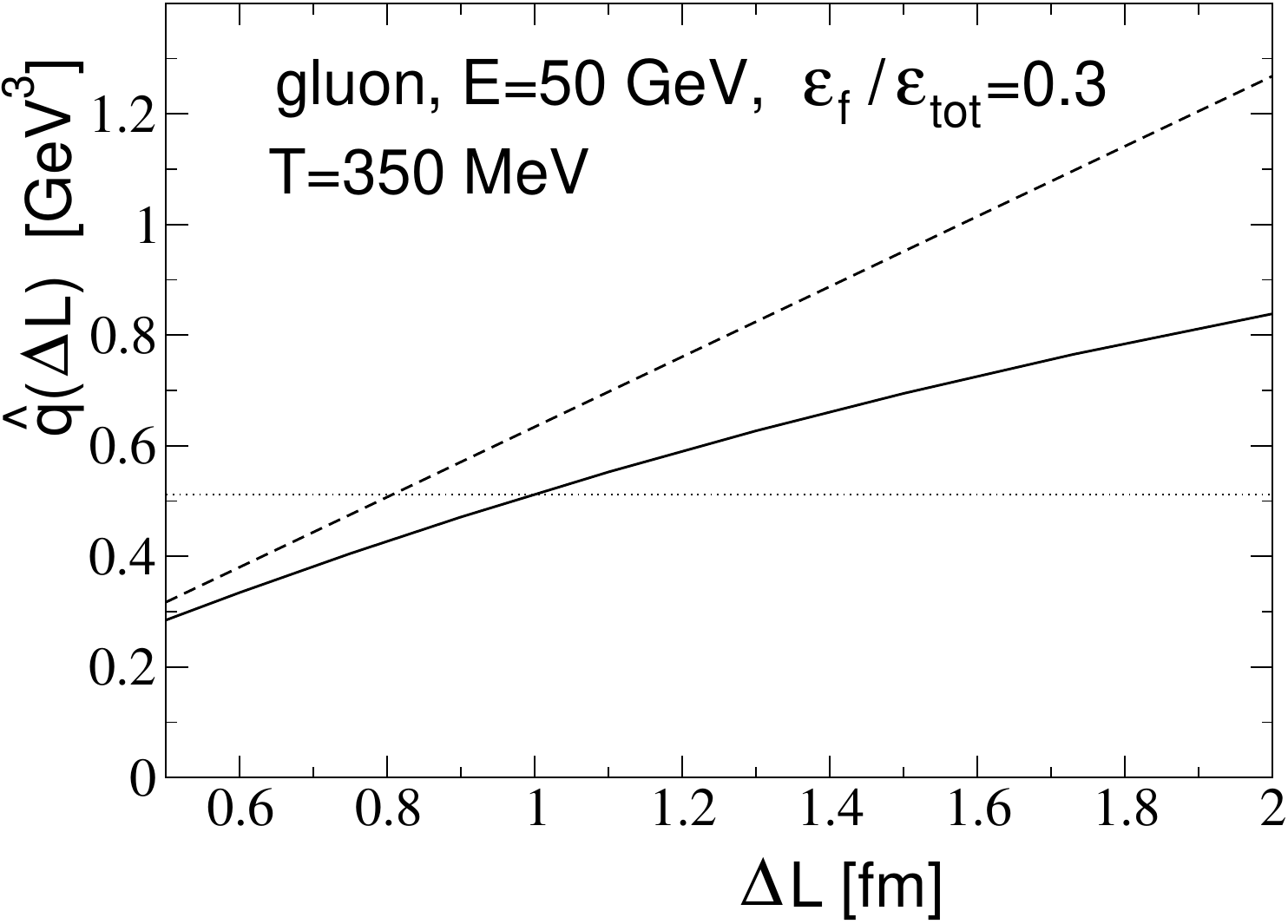}  
\includegraphics[height=3.7cm,angle=0]{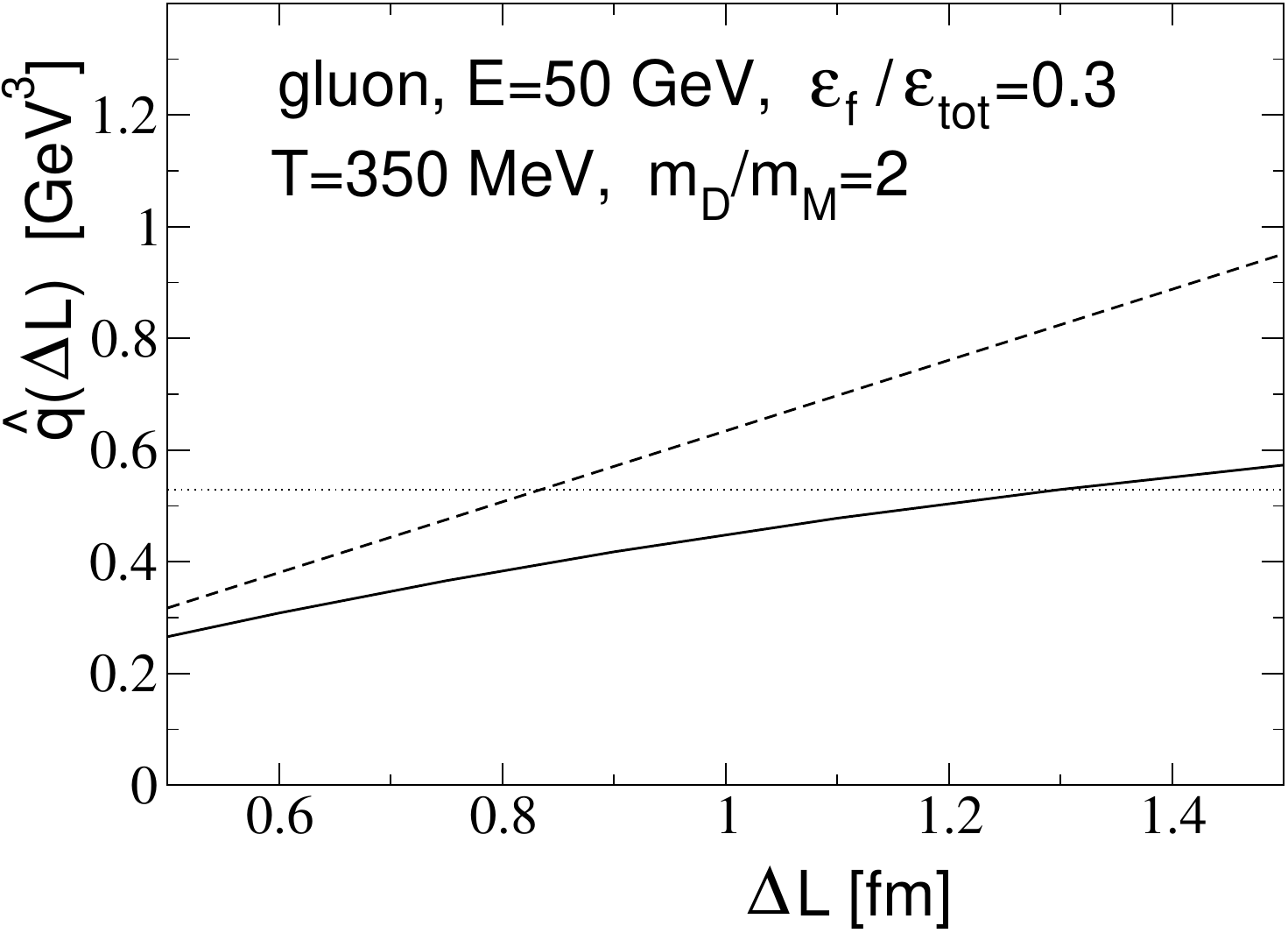}  
\includegraphics[height=3.7cm,angle=0]{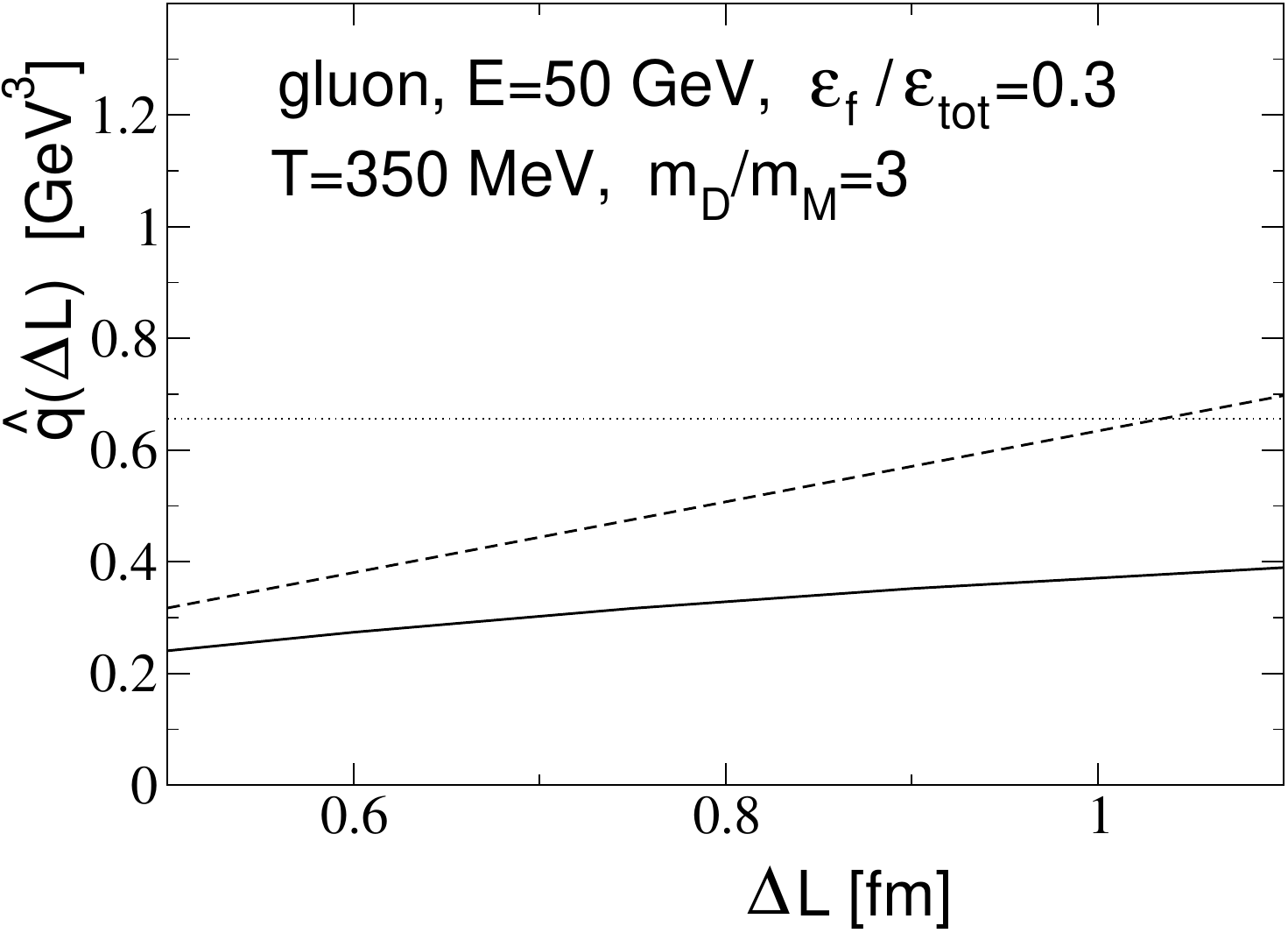}  
\end{center}
\caption[.]{
The same as in Fig. 5 for gluon.
}
\end{figure}

In this section, we present numerical results for $\hat{q}$
using formulas of previous section\footnote{Note that our numerical calculations
  demonstrated that for gluons the last two states made of color neutral
  gluons $AB$ and $BA$ do not contribute to $\hat{q}$, i.e., for
  gluons calculations can be performed with $8\times 8$ matrix.}.
We use 
the QCD coupling constant $g=2$.
To characterize the magnitude of the background color field
we use the ratio of the chromomagnetic field energy density $\epsilon_f=B^2/2$
to the total QGP energy density
$\alpha=\epsilon_f/\epsilon_{tot}$.
We define the total QGP energy density
in terms of the equilibrium temperature in the ideal gas model: 
$\epsilon_{tot}=(\frac{8\pi^2}{15}+\frac{7\pi^2N_f}{20})T^4\approx 13.9T^4$
(we take $N_f=2.5$). We assume that the magnetic field is isotropic.
In this case for the Gaussian parameter $\sigma$ in (\ref{eq:550})
we have $\sigma^2=(2/3)\alpha \epsilon_{tot}$.
We perform calculations of $\hat{q}$ for $T=350$ MeV, 
which seems to be reasonable for the initial stage of the QGP
evolution at $\tau\sim 1-2$ fm for heavy ion collisions at the LHC energies.
In the absence of realistic calculations
of the turbulent QCD matter evolution in $AA$ collisions, the characteristics
of the turbulent magnetic fields can not be obtained theoretically.
We perform calculations for for $\alpha=0.2$ and $0.3$.
Such values of $\alpha$
can lead to reasonable values of the ratio $\eta/s$
in the scenario with the dominant contribution of the turbulent
fields to the QGP shear viscosity
\cite{Asakawa1,BMuller1,Muller-Wang-qhat}.
The ratio of the chromomagnetic energy to the total QGP energy
substantially higher than $0.2-0.3$ looks unrealistic.

For quarks we perform calculations of $\hat{q}$ for $\Delta L$ in the interval
between 0.5 to 2 fm. We present the results obtained for the number of
slices  (in Eq. (\ref{eq:570})) $M=15$. For the HTL version with $m_D/m_M=3$,
that has the smallest $\lambda_{mfp}$, the estimated errors are $\lsim 2$\%
at $L\sim 2$ fm, and are $\lsim 0.5$\% at $L\lsim 1$ fm (for
$m_D/m_M=2$ the errors becomes smaller by a factor of $\sim 2-3$).
For the static model even at $L\sim 2$ fm the errors are
$\lsim 0.3$\%.
For gluons the needed computational resources are considerably
larger than for quarks (for the same value of $M$). For this
reason we performed calculations for $M=10$. For the HTL
version we restricted the maximal values of $\Delta L$ ($\Delta L<1.1(1.5)$ fm
for $m_D/m_M=3(2)$) to avoid regions where the errors may be too large.
For the static model we present the results for the interval $0.5<\Delta L<2$
fm (in this interval the estimated errors are $\lsim 3$\%
at $L\sim 2$ fm, and are $\lsim 0.7$\% at $L\lsim 1$ fm).

In Figs. 4--7 we present results for the contribution
to $\hat{q}$ for quark and gluon from scattering on the
thermal constituents, and from
scattering in the random chromomagnetic fields without and with the effect
of the parton color randomization. The calculations are performed
for the parton energy $E=50$ GeV. Note that in our model only the
thermal contribution to $\hat{q}$ depends on the parton energy
(due to the energy dependence of the Coulomb logarithm).
From Figs. 4--7 one can see that the effect of the parton color randomization
reduces the turbulent contribution to $\hat{q}$ 
by a factor of $\sim 0.8$ at $\Delta L\sim 1$
and $\sim 0.65$  at $\Delta L\sim 2$ fm for
the static model (the reduction is approximately the same
for quarks and gluons).
For the HTL scheme
the effect is stronger. For
$m_D/m_M=2(3)$,  the color randomization reduces the
turbulent $\hat{q}$ for quarks by a factor of $0.7(0.6)$ at $\Delta L\sim 1$ fm
and by a factor of $0.5(0.4)$ at $\Delta L\sim 2$ fm. 
The magnitude of the
reduction of the turbulent $\hat{q}$ for gluons
due to the color randomization at $\Delta L\sim 1$ fm in the HTL scheme
is similar to that for quarks.

Note that, since the turbulent contribution to $\hat{q}$ is approximately
energy independent,
one can expect that our results should be valid for the
thermal partons as well.
The effect of the parton color randomization
should lead to some increase of the turbulent
shear viscosity as compared to predictions without
the color randomization (like that of \cite{Asakawa1,BMuller1}).
For the HTL scheme with $m_D/m_M=2(3)$ the ratio $\eta/s$
should become bigger by a factor of $\sim 1.2-1.5(1.3-1.8)$ for
$\Delta L\sim 0.5-1$ fm.

Finally, it is worth noting that the suppression of $\hat{q}_f$
due to the color randomization of fast partons may be 
enhanced by the effects of the running coupling, that have been ignored
in the present analysis. The lattice calculations of \cite{Bazavov_al1}
show that at small virtualities the in-medium $\alpha_s$ may grow
up to $\sim 0.5-0.8$. The growth of $\alpha_s$ at low
momenta (and the perturbative logarithmic decrease at high momenta) is
more important for
the parton color randomization length
$\lambda_{mfp}$ entering (\ref{eq:650}) than for
the thermal transport coefficient $\hat{q}_{th}$ (for which the relative
contribution of the low momentum region is suppressed by the factor $q^2$).
For this reason,
the suppression effect of the color randomization on the turbulent
contribution to $\hat{q}$ should be bigger for the 
scheme with the running coupling (if the models with the fixed and
the running coupling lead to similar predictions for the thermal transport
coefficient).

%%%%%%%%%%%%%%%%%%%%%%%%%%%%%%%%%%%%%%%%%%%%%

\section{CONCLUSIONS}
We  have  analyzed  the  effect of the parton color randomization
on the $p_T$ broadening  of fast  partons  in  the QGP with
turbulent color fields that can be generated in the QGP formed
in $AA$ collisions due to the non-abelian Weibel instabilities.
Calculations of the transverse
momentum broadening  of a fast parton
traversing the turbulent QGP ignoring its interaction with
the QGP constituents give the transport coefficient $\hat{q}$
approximately proportional to the product of the mean field energy
$\epsilon_f$ and the correlation length $r_c$ of the turbulent color fields.
The gluon exchanges between  the fast parton and the thermal partons
change the color charge of the fast parton, that lead to
random variation of the Lorentz force experienced by the fast parton.
This acts as a reduction of the correlation length of the
turbulent color fields, and reduces the transport coefficient. 

We performed calculations of $\hat{q}$ for a simplified model
of fluctuating color fields in the form of
alternating sequential transverse layers of thickness $\Delta L$
of homogeneous transverse
chromomagnetic fields with random orientation in the SU(3) group, and
gaussian distribution in the magnitude.
We demonstrated that calculation of $\hat{q}$ may be reduced to
calculation of $p_T$ broadening in a single layer.
We presented the one-layer $\langle p_T^2 \rangle$ in a form
which is convenient for numerical
simulations.
We performed calculations using
for the color exchanges between the fast parton and the thermal partons
the static model of the Debye screened
color centers  \cite{GW} and the HTL scheme \cite{HTL} with a nonzero
magnetic mass \cite{AGZ}.  
Our numerical results show that the color randomization
can lead to a sizable reduction of the turbulent contribution
to $\hat{q}$. We find that the reduction of $\hat{q}$ due to
the color randomization is bigger for the HTL scheme, for which
the effect grows with increasing $m_D/m_M$. 
For the HTL version with $m_D/m_M=3$ we obtained the suppression of
the turbulent
contribution to $\hat{q}$
by a factor of $\sim 0.6$ for $\Delta L\sim 1$
and $\sim 0.4$  for $\sim 2$ fm.
We have found that the reduction of the turbulent
contribution to $\hat{q}$ due to the parton color randomization
is very similar for quarks and gluons.
The magnitude of the suppression is weakly dependent
on the average energy of the turbulent color fields in the QGP.
%%%%%%%%%%%%%%%%%%%%%%%%%%%%%%%%%%%%%%%%

\section*{Acknowledgments}
This work is supported by the State program 0033-2019-0005.

\end{document}